\documentclass[epj]{svjour}

\usepackage{amsmath}
\usepackage{graphics}
\usepackage{graphicx}
\usepackage{dcolumn}
\usepackage{bm,color}
\usepackage{footnote}
\usepackage[utf8x]{inputenc}
\usepackage[T1]{fontenc}
\usepackage{mathptmx,comment}
\begin{document}
\title{Mechanics and Thermodynamics of a New Minimal Model of the Atmosphere} 
\author{Gabriele Vissio\inst{1,}\inst{2,} 
\and Valerio Lucarini\inst{3,}\inst{4}
\thanks{Corresponding author. \emph{Contact Email:} v.lucarini@reading.ac.uk}%
}                     
%
%
\institute{CEN, Meteorological Institute, University of Hamburg, Hamburg, Germany \and Institute of Geosciences and Earth Resources (IGG) - National Research Council (CNR), Torino, Italy \and Department of Mathematics and Statistics, University of Reading, Reading, UK \and Centre for the Mathematics of Planet Earth, University of Reading, Reading, UK}
\date{Received: \today}
%
\abstract{The understanding of the fundamental properties of the climate system has long benefitted from the use of  simple numerical models able to parsimoniously represent the essential ingredients of its processes. Here we introduce a new  model for the atmosphere  that is constructed by supplementing the now-classic Lorenz '96 one-dimensional lattice model with temperature-like variables. The model features an energy cycle that allows for conversion between the kinetic and potential forms and for introducing a notion of efficiency. The model's evolution is controlled by two contributions - a quasi-symplectic and a gradient one, which resemble (yet not conforming to) a metriplectic structure. After investigating the linear stability of the symmetric fixed point, we perform a systematic parametric investigation that allows us to define regions in the parameters space where at steady state stationary, quasi-periodic, and chaotic motions are realised, and study how the terms responsible for defining the energy budget of the system depend on the external forcing injecting energy in the kinetic and in the potential energy reservoirs. Finally, we find preliminary evidence that the model features extensive chaos. We also introduce a more complex version of the model that is able to accommodate for multiscale dynamics and that features an energy cycle that more closely mimics the one of the Earth's atmosphere.}

%
      
      \maketitle
      
      \section{Introduction} \label{intro}
The climate is a non-equilibrium system whose dynamics is primarily driven by the uneven absorption of solar radiation, which is mainly absorbed near the surface and in the tropical latitudes, rather than aloft and in the mid-high latitudes, respectively. The system reacts to such an inhomogeneity in the local energy input through a complex set of instabilities and feedbacks affecting its dynamical processes and thermodynamic and radiative fluxes. Such processes lead to an overall reduction in the temperature gradients inside the the system and allow for the establishment of approximate steady state conditions \cite{Peixoto1992,Lucarini2014}.  

An example of the re-equilibration mechanism can be described as follows.  The large scale energy transport is mainly performed by atmospheric  disturbances in the form of synoptic and (to a lesser extent) planetary eddies which are, in turn, fuelled by the baroclinic conversion of available potential energy into kinetic energy, which is then dissipated by friction. In turn, the presence of large low-high latitudes temperature gradients is responsible for the presence of a reservoir of available potential energy, which is continuously replenished thanks to the dishomogeneity of the radiative energy budget across the globe \cite{Peixoto1992}.  This is the core of the celebrated Lorenz energy cycle, which provides a powerful representation of the climate as an engine \cite{Lorenz1967}. The thermodynamic viewpoint on the climate allows to define its efficiency and irreversibility  \cite{Pauluis2,Lucarini2009,Ambaum2010,Lucarini2010b,Laliberte2015,Lembo2019}.

The reconstruction, interpretation, and analysis of observative data (now facilitated by the recent advances in data science); analytical tools borrowed (mostly) from mathematics and physics; and numerical simulations all contribute to our understanding of the climate. This task is exceedingly demanding since the system features non-trivial variability on a vast range of temporal and spatial scales and, furthermore, our ability to observe it has changed enormously over time. Additionally, the presence of periodic as well as irregular fluctuations in the boundary conditions do not allow the climate to reach an exact steady state \cite{Ghil2015,GhilLucarini2020}.  

 One of the features of the numerical investigation of the climate system is the reliance on hierarchies of models. In other terms, climate phenomena are investigated using a full range of models going from low dimensional ones to state-of-the-art Earth system models, able to represent with higher precision many aspects of the climate. See a discussion of the meaning and of the use of hierarchies of climate models in \cite{Schneider1974,Held2005,Lucarini2013,GhilLucarini2020}. It is important to remark that, given the multiscale nature of the climate system, the heterogeneity of its subdomains, and the number of the active physical, chemical, and biological processes, the endeavour of construcing a model able to directly simulate all of them appears as a Sisyphean task, whilst, instead, the parametrization of the effect of the unresolved scales on those that are explicitly simulated is an essential component of any reasonable model of the Earth system\cite{Palmer2008,Franzke2015,Berner2017}.

Low-order models have played and still play today a very important role for improving our understanding of the geophysical flows. Apart from the landmark 3-dimensional model developed by Lorenz in 1963  \cite{Lorenz1963} starting from the truncation of the equations describing the Rayleigh-Benard convection introduced by Saltzman \cite{Saltzman1962}, simple models have been key to scientific advances in oceanography  \cite{Stommel1961,Veronis1963,Rooth1982}, dynamical meteorology \cite{Charney1979,Lorenz1984,Lorenz1996}, climate dynamics \cite{Budyko1969,Sellers1976,Ghil1976,Fraedrich1979}, turbulence  \cite{Gledzer1973,Biferale1995,Lvov1998,Biferale2003} and convection \cite{Brandenburg1992,Mingshun1997}, among others. Additionally, low-order models supplemented by stochastic forcings have also provided the backbone of the stochastic theory of climate \cite{Hasselmann1976,Saltzman2001,Imkeller2001}. Aside from sheer mathematical-related curiosity, many of these models were created in order to shed some light on specific problems by using a physically-meaningful benchmark tool that is easier to analyse mathematically and faster to simulate numerically with respect to more complex models.


\subsection{The Lorenz '96 model} \label{L96 model}

Of special relevance for the present study is the now-celebrated Lorenz '96 model \cite{Lorenz1996,Lorenz2005} (hereafter L96), whose structure is briefly recapitulated below. The model consists of a lattice of $N$ gridpoints, whose state is described by a variable. The model has periodic boundary conditions, so that it can loosely be interpreted as describing the properties of the atmosphere along a latitudinal circle. The model features in an extremely simplified - almost metaphorical - way the main processes of the atmosphere: forcing, dissipation, and advection. Two versions of the model have been proposed: a one-level version, where the dynamics takes place on a single length scale, and a two-level version, where the lattice is augmented in order to describe dynamical processes on smaller spatial and temporal scales. The two-level version of the L96 model has the especially attractive property that the time-scale separation between the fast and the slow variables can be  controlled by modulating one parameter. 

The L96 model has rapidly gained relevance among geoscientists, physicists, and applied mathematicians, as it has become a benchmark testbed for parametrizations \cite{Orrell2003,Wilks2005,Abramov2016,Arnold2013,Vissio2018,Vissio2018b,Chattopadhyay2020}, for studying extreme events \cite{Blender2013,Bodai2017,Sterk2017,Hu2019}, for developing data assimilation schemes \cite{Trevisan2004,Trevisan2010,Hu2017,Brajard2020}, {\color{black}for developing ensemble forecasting techniques \cite{Wilks2006,Duan2016,Lecarrer2020},} for studying the properties of Lyapunov exponents and covariant Lyapunov vectors \cite{Pazo2008,Hallerberg2010,Karimi2010,Carlu2019}, for developing and testing ideas in nonequilibrium statistical  mechanics \cite{AbramovM2008,Lucarini2011,Lucarini2012,Gallavotti2014,Abramov2017}, and for investigating bifurcations \cite{Ott1993,Broer2002,Orrell2003a,vanKekem2018PhysD,vanKekem2018NPG,vanKekem2019,Kerin2020}. By looking at these references, the reader can find  a very thorough analysis of the properties of the L96 model. 

The one-level L96 model can be written as:
\begin{equation} \label{lor1a}
  \frac{dX_k}{dt}=X_{k-1}(X_{k+1}-X_{k-2})-\gamma X_k+F,
\end{equation}
with boundary conditions:
\begin{equation}
  X_{k-N}=X_{k+N}=X_k.
\end{equation}
$k=1,\ldots,N$ is the index of the gridpoints defining the lattice, the nonlinear term defines a nontrivial process of advection, $F$ is an external forcing, and $\gamma$ (usually taken with unitary value) modulates the dissipation. In the unforced and inviscid regime - i.e. setting $F=\gamma=0$, the energy of the system, expressed as the sum of the squares of the variables, is conserved:
\begin{equation} \label{conslor96}
  \frac{dE}{dt}=\frac{d}{dt}\sum_{k=1}^{N}\frac{X_k^2}{2}=\sum_{k=1}^{N}X_k\frac{dX_k}{dt}=0.
\end{equation}
If $\gamma=1$ and $N\gg1$, the model's attractor is a fixed point for $0\leq F\leq8/9$. As $F$ is increased, the fixed point $X_1=X_2=\ldots = X_K = F$ loses stability as the system undergoes bifurcations leading to a quasi-periodic behavior for moderate values of $F$ and chaotic behaviour for {\color{black}$F\geq 5.0$} \cite{Lorenz2005}. {\color{black}This is only a rough description of the complexity of the bifurcations taking place in the L96 model as $F$ is changed: as discussed in detail in \cite{vanKekem2018PhysD,vanKekem2018NPG,vanKekem2019}, the properties of the system depend on $N$ in a very nontrivial way in the regime of moderate forcing. In the regime of strong forcing and developed turbulence, instead, some sort of universality emerges, as the L96 model is strongly chaotic }when $F\geq 8$ and its properties are extensive with respect the number of nodes $N$ \cite{Gallavotti2014}. By and large, the mechanism of instability of the L96 model boils down to exchanges of energy between the symmetric state  and the perturbations away from it, as particularly clear in the case of the linear instability analysis \cite{Lorenz2005}. Nonetheless, the L96 model clearly features only one form of energy, which we may refer to as \textit{kinetic}. 

\subsection{This Paper}
In this paper we propose an extension of the L96 model whereby a second variable is attached to each gridpoint, representing, metaphorically, the local thermodynamical properties, which are advected by the dynamical variables of the L96 model, and undergo forcing and diffusion. The model proposed here features a meaningful definition of energy that includes the kinetic part already present in the L96 model plus a potential part associated to the fluctuations of the \textit{temperature} in the domain. The fundamental advantage of the new model proposed here and analysed in detail in Sect. \ref{One-layer model} is that it features an energy cycle that allows for conversion between the kinetic and potential forms of energy. Conceptually, this mirrors the change one has going from one a one-layer quasi-geostrophic model, where only barotropic processes are possible, to a two-layer quasi-geostrophic model, which, instead, features a coupling between dynamical and thermodynamic processes via baroclinic conversion \cite{Holton2004}. Note that the Lorenz '63 model \cite{Lorenz1963} and, more completely so, its extensions to higher order modes \cite{Chen2006} features a nontrivial energetics where exchanges take place between potential and kinetic energy \cite{LucariniFraedrich2009}. The energy of the system defines the symplectic component that contributes - together with the metric one - to defining the evolution of the model \cite{Blender2013}.

The rest of the paper is structured as follows. Section \ref{One-layer model} provides a thorough introduction to the one-level version of the model presented here. The evolution equations are presented together with the rationale on which the model is based. Additionally, a detailed analysis of its mechanical and thermodynamic properties is carried out. Finally, the linear stability analysis of the symmetric fixed point is also described. In Sect. \ref{Tests} we present the results of a large set of numerical integrations of the model, aimed at exploring its properties in a rather vast range of values of two main parameters, which control the input of energy in the kinetic and potential form. We discuss the thermodynamics of the model in terms of mean values and the fluctuations of the main terms describing the energetics of the model. Such a physical characterisation of the model is complemented by the analysis of how the first Lyapunov exponent depends on the two considered parameters, in order to be able to separate  the regions where the asymptotic dynamics of the system takes place in a regular vs in a strange attractor \cite{Eckmann1985,Ott1993} We will discover a non-trivial interplay between the two sources of forcings applied to the model. We then perform a preliminary analysis for assessing to what extent the system obeys extensive chaos. In Sect. \ref{Conclusions} we summarise the main features of the model and the results obtained so far, and propose future lines of investigations. 
 As in the case of original L96 model, the model introduced here can be formulated in a two-level fashion - see Appendix \ref{App_NM} -, with non trivial couplings among different levels and variables and with a fairly sophisticated energetics, which is conceptually rather similar to the one described by the Lorenz energy cycle in the atmosphere. The analysis of the properties of the two-level model will not be performed in this paper and will be the subject of future studies.

\section{Model Formulation and Properties} \label{One-layer model}


We want to extend the standard L96 model presented in the introduction by adding a second set of variables for all the gridpoints $k=1,\ldots,N$. The goal is to construct a toy model able to describe in a very simple yet conceptually correct way the interaction between dynamical and thermodynamical processes of the atmosphere. The evolution equations of the model we propose in this contribution are the following: 

\begin{equation} \label{eq:model1a}
  \frac{dX_k}{dt}=X_{k-1}(X_{k+1}-X_{k-2})-\alpha\theta_k-\gamma X_k+F,
\end{equation}
\begin{equation} \label{eq:model1b}
  \frac{d\theta_k}{dt}=X_{k+1}\theta_{k+2}-X_{k-1}\theta_{k-2}+\alpha X_k-\gamma\theta_k+G,
\end{equation}
with $k=1,\ldots,N$. The variable $\theta_k$ can be loosely interpreted as temperature at the grid-point $k$. The boundary conditions are defined as 
\begin{equation}
\begin{split}
  X_{k-N}=X_{k+N}=X_k,\\
  \theta_{k-N}=\theta_{k+N}=\theta_k,\\
\end{split}
\end{equation}
The variable  $\theta_k$ undergoes a constant forcing $G$, a linear dissipation term, and a nonlinear term representing, loosely speaking, the advection performed by the $X$ variables. Additionally, $\theta_k$ and $X_k$ are linearly coupled through a term proportional to $\alpha$. The purpose of this coupling is to represent, in a very simplified way, the effect of correlated thermal and dynamical fluctuations on the dynamics, which allow for an exchange between kinetic and potential energy associated with thermal fluctuations, as discussed below. The introduction of a term proportional to $\alpha$ is the only - yet important - modification in the dynamics of the $X$ variables for this model as compared to the classical L96 model, see Eq. \ref{lor1a}. In what follows, we consider $F,G,\alpha,\gamma \geq0$. 

The coupling between the $X$ and the $\theta$ variables  is constructed in such a way that in the unforced and inviscid limit ($F=G=\gamma=0$) the total energy  of the system $$E=K+P=\sum_{k=1}^{N}(\frac{X_k^2}{2}+\frac{\theta_k^2}{2})$$ 
given by the sum of its kinetic and potential components, is conserved:
\begin{equation} \label{cons}
	\frac{dE}{dt}=\frac{dK}{dt}+\frac{dP}{dt}=\sum_{k=1}^{N}X_k\frac{dX_k}{dt}+\sum_{k=1}^{N}\theta_k\frac{d\theta_k}{dt}=0,
\end{equation}
whereas, in general, $K$ and $P$ are not separately conserved. {\color{black}The quadratic functional form of the potential energy is inspired by the fact that the available potential energy in the global circulation of the atmosphere is approximately proportional to the variance of the temperature fluctuations \cite{Lorenz1967,Peixoto1992,Grotjahn1993}}. The dynamical role of the function $E$ is explored in the next section.





\subsection{Mechanics}

By definition, the time derivative of any smooth observable $\Psi(X_1,\ldots,X_K,\theta_1,\ldots,\theta_k)$ is obtained by applying the generator of the Koopman operator to $\Psi$, as follows:
$$\frac{\mathrm{d}\Psi}{\mathrm{d}t}=\mathcal{L}[\Psi].$$
We will now show that linear operator $\mathcal{L}$  can be written as the sum of a contribution coming from a symplectic (indeed, {\color{black}quasi-symplectic}, for the reasons detailed at the end of this section) term and a contribution coming from a gradient term. 
Indeed, we can write:
\begin{equation}\label{koopman}
\frac{\mathrm{d}\Psi}{\mathrm{d}t} =\mathcal{L}[\Psi]=\left\{\Psi,E\right\}+\langle \Psi,\Gamma\rangle
\end{equation}
where  $\left\{A,B\right\}=-\left\{B,A\right\}$ is a suitably defined Poisson bracket for the functions $A$ and {\color{black}$B$}, while $\langle A,B\rangle=\langle B,A\rangle$ gives the gradient contribution. The evolution equations \ref{eq:model1a}-\ref{eq:model1b} are
obtained by setting $\Psi = X_i$ and $\Psi=\theta_i$, $i=1,\ldots,N$, respectively.

We have that $\langle A,B\rangle=\left(\partial_{X_i} A\right) \left(\partial_{X_i} B\right) +\left(\partial_{\theta_i} A\right) \left(\partial_{\theta_i} B\right)$, where we use the Einstein convention for the indices.  The function $\Gamma$ defining the gradient contribution to the dynamics is:
\begin{equation}\label{gamma}
\Gamma=-\gamma E + F \sum_{k=1}^N X_k + G \sum_{k=1}^N \theta_k=-\gamma E +F \Xi + G \Theta, 
\end{equation}
where $\Xi=\sum_{k=1}^N X_k $ and $\Theta =\sum_{k=1}^N \theta_k$. it is clear that such a component describes the irreversible dynamics as it vanishes in the unforced, inviscid limit $F=G=\gamma=0$. 

We then discuss the symplectic term associated with the Poisson bracket. We have that 
\begin{equation}\label{symple}
\begin{split}
\left\{A,B\right\}&=\left(\partial_{X_i} A\right) J_{ij} \left(\partial_{X_j} B\right)+ \left(\partial_{\theta_i} A\right) Y_{ij} \left(\partial_{\theta_j} B\right)\\
&+\left(\partial_{X_i} A\right) L_{ij} \left(\partial_{\theta_j} B\right)+\left(\partial_{\theta_i} A\right) M_{ij} \left(\partial_{X_j} B\right)
\end{split}
\end{equation}
where
\begin{equation}\label{poisson}
\begin{split}
J_{ij}&=X_{i-1}\delta_{i+1,j}-X_{j-1}\delta_{i,j+1}\\
Y_{ij}&=X_{i+1}\delta_{i+2,j}-X_{i-1}\delta_{i-2,j}\\
L_{ij}&=-\alpha \delta_{i,j}\\
M_{ij}&=\alpha \delta_{i,j}\\
\end{split}
\end{equation}
It is clear that the energy $E$ is the generator of time translations according to the symplectic contribution and the antisymmetry of the Poisson brackets enforces the corresponding  conservation law already discussed in Eq. \ref{cons}. 

We remark that, in the {\color{black}inviscid} and unforced limit the system is not Hamiltonian because the Poisson brackets do not fulfill the Jacobi identity $\{A,\{B,C\}\}+\{C,\{A,B\}\}+\{B,\{C,A\}\}=0$. Because of this and of the fact that $\{\Gamma,E\}\neq 0$, the system given in Eqs. \ref{eq:model1a}-\ref{eq:model1b} is not metriplectic, i.e. the standard generalisation of Hamiltonian system to the dissipative case. \cite{Kaufman1984}. Note that the dynamics of dissipative fluids is, instead, metriplectic \cite{Grmela1984}, and so is the dynamics of the (extended) Lorenz '63 model, which is in fact derived from the Rayleigh-B\'enard equations through systematic modal truncation \cite{Blender2013}. The lack of an underlying Hamiltonian skeleton confirms the well-known fact that the L96 model cannot be easily related to any model of fluid flows.

\subsection{Thermodynamics}

Using Eqs. \ref{koopman}-\ref{gamma} we obtain the time evolution of the energy of the system:
$$
\frac{\mathrm{d}E}{\mathrm{d}t}=\langle \Psi,\Gamma\rangle=-2\gamma E + F \Xi  + G \Theta = \Gamma-{\color{black}\gamma E}
$$
which implies that, at steady state $2\gamma \bar E = F\bar \Xi + G \bar \Theta$, where $\bar \Psi$ is the long term average of the quantity $\Psi$.

\begin{figure}
\includegraphics[width=0.6\linewidth]{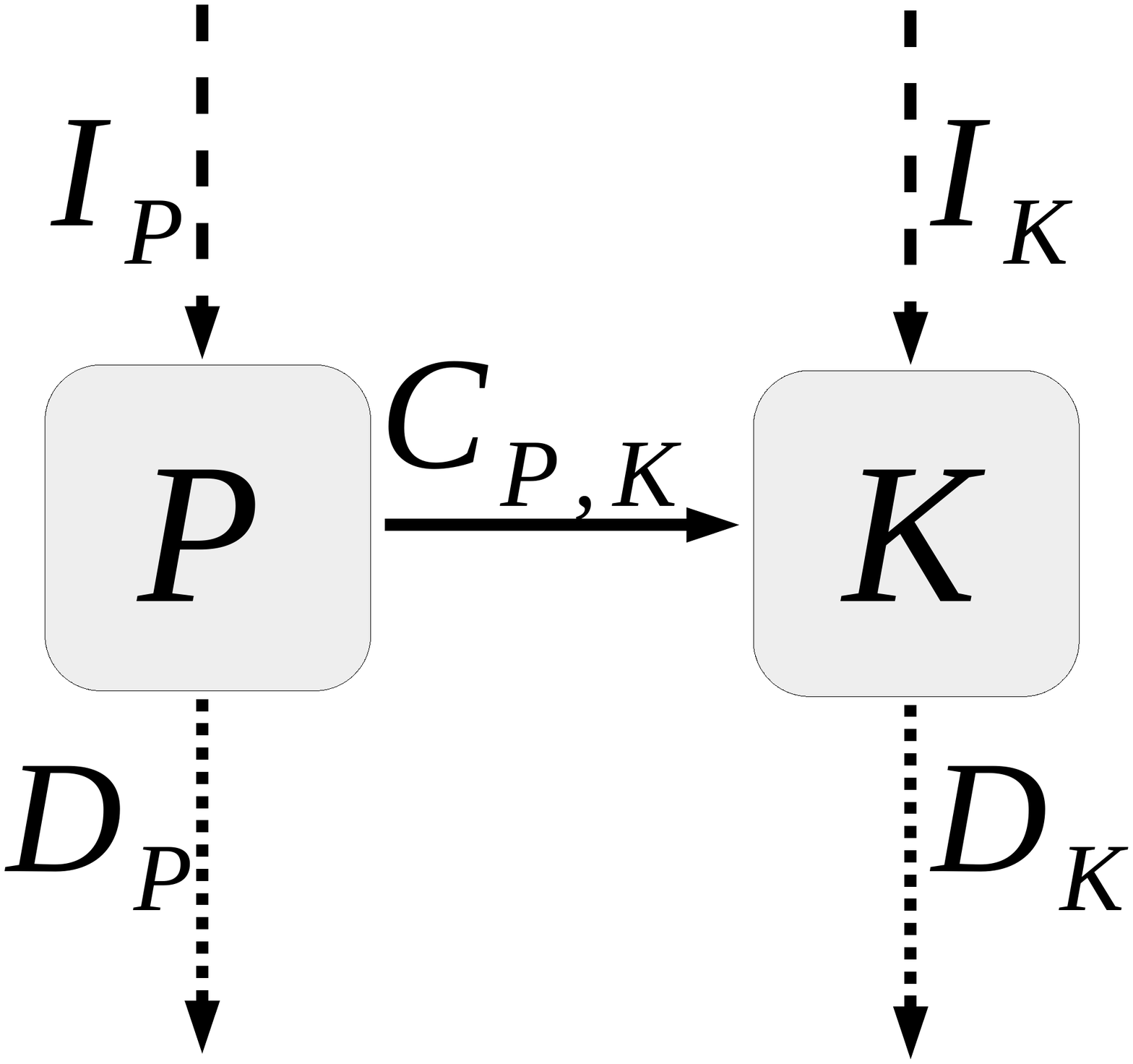}
\caption{\label{BnP_1-layer} Energetics of the model presented in Eq. \ref{eq:model1a}-\ref{eq:model1b}. 
We indicate the fluxes of energy in and out of the reservoirs of kinetic (K) and potential (P) energy. Dashed lines represent input of energy; dotted lines represent energy dissipation; and solid lines represent energy conversion terms. {\color{black}The direction of the arrows indicates a positive energy flux.} See text for details.} 
\end{figure}

We next analyse the separate budget of the kinetic and potential energy. By inserting $K$ and $P$ in Eq. \ref{koopman} we obtain: 

\begin{equation} \label{LEC1}
	\frac{\mathrm{d}K}{\mathrm{d}t}  =\sum\limits_{k=1}^N X_k\frac{\mathrm{d}X_k}{\mathrm{d}t}=I_K-D_K+C_{P,K}\quad I_K=F \Xi,\quad D_K=2\gamma K,\quad C_{P,K}=-\sum\limits_{k=1}^N \left(\alpha X_k \theta_k \right),
\end{equation}
\begin{equation} \label{LEC2}
	\frac{\mathrm{d}P}{\mathrm{d}t} =\sum\limits_{k=1}^N \theta_k\frac{\mathrm{d}\theta_k}{\mathrm{d}t}=I_P-D_P-C_{P,K}\quad I_P=G \Theta,\quad D_P=2\gamma P,\\
\end{equation}
where $I_K$ ($I_P$) is the rate of input of kinetic (potential) energy, $D_K$ ($D_P$) is the dissipation rate of kinetic (potential) energy, and $C_{P,K}$ the conversion rate from potential to kinetic energy. {\color{black} We remark that the input and  dissipation of energy in either kinetic or potential form is due by the metric component of the dynamics. Instead, the conversion of energy between the potential and kinetic form is controlled by the Poisson brackets given in Eqs. \ref{symple}-\ref{poisson}. Nonetheless, the components $J$ and $Y$ of the Poisson brackets}, which describe advection, do not give any net contribution. Equations \ref{LEC1}-\ref{LEC2} describe the  energetics of the model presented in this work, which is represented by the diagram shown in Fig. \ref{BnP_1-layer}. One can draw a parallel between the energetics of this model and the Lorenz energy cycle of the atmosphere, where, as well known, the input of energy comes almost entirely through the potential energy channel via baroclinic forcing associated with the differential heating of low versus high latitude regions \cite{Lorenz1967,Peixoto1992}. The two-level version of the model introduced in this paper features an energetics that is conceptually closer to the one of the true atmosphere because it is able to describe energy cascades across scales on top of energy conversion processes, see Appendix \ref{App_NM}.

At steady state conditions one has $2\gamma \bar{K} = F \bar\Xi + \bar{C}$ and $2\gamma \bar{P} = G \bar\Theta -\bar{C}$, which relate the size of the reservoirs of kinetic and potential energy to intensity of the acting forcings and energy exchange. We can also introduce a notion of efficiency of this model $\eta=\bar{C}/(F\bar{\Xi}+G\bar{\Theta})=\bar{C}/(2\gamma \bar{E})$, which relates the amount of energy exchanged between the two reservoirs of energy to the total energy input. Since $X_k^2+\theta_k^2\geq2|X_k\theta_k|$ $\forall k$, we have that $2 E\geq |C|/\alpha$. Therefore, $|\eta|\leq\alpha/\gamma$, which provides a constraint on the efficiency of the system. Note that $\eta$ is positive if, on the average, energy is converted from potential to kinetic, and negative otherwise.

Note that $K\geq 0$ and $P\geq0$ by definition. As a result, if $G,\gamma> 0$ and $F=0$, one has $\bar C\geq 0$ (on the average we have an energy flux from potential to kinetic)\footnote{Note that this is, to a very good approximation, what applies to the climate system as a whole, because the geophysical fluids do not receive any input of mechanical energy, apart from the very small lunar and solar tidal forcing.}, whereas if $F,\gamma> 0$ and $G=0$, $\bar C\leq 0$ (on the average we have an energy flux from kinetic to potential). If $F=G=0$, instead, we have that $\bar{K}=\bar{P}=\bar{E}=0$, where the origin is a stable fixed point for the system.

\subsection{Linear Stability Analysis}\label{linearstability}
We investigate the linear stability of the system analysed here around the fixed point corresponding to the symmetric solution $X_j=X_k=const$, $1,\ldots,j, k,\ldots N$ and $\theta_j=\theta_k=const$, $1,\ldots,j, k,\ldots N$ . By plugging this ansatz in  Eqs. \ref{eq:model1a}-\ref{eq:model1b} one gets:
\begin{equation} \label{fixedpointa}
  X_k=\tilde X=\frac{\gamma F-\alpha G}{\gamma^2+\alpha^2} \forall k,
\end{equation}
\begin{equation} \label{fixedpointb}
    \theta_k=\tilde \theta=\frac{\alpha F+ \gamma G}{\gamma^2+\alpha^2} \forall k.
\end{equation} 
Taking inspiration from \cite{Lorenz2005}\footnote{{\color{black}See also the rather sophisticated analysis of the stability of the L96 model presented in \cite{vanKekem2018NPG,vanKekem2018PhysD,vanKekem2019}.}}, we then investigate the linear stability of this solution  by substituting $X_k=\tilde X+A\exp(\sigma t)\exp(\mathrm{i}\kappa k-\mathrm{i}\omega t)$ and $\theta_k=\tilde \theta+B\exp(\sigma t)\exp(\mathrm{i}\kappa k-\mathrm{i}\omega t)$ in Eqs. \ref{eq:model1a}-\ref{eq:model1b}, {\color{black}where $\tilde X$ and $\tilde \theta$ have been defined in Eqs. \ref{fixedpointa} and \ref{fixedpointb}, respectively;} $A$ and $B$ are complex constants, $\sigma$ is a real number defining the growth rate (if positive) of the  amplitude of the wave, whilst $\kappa$ is the wavenumber and $\omega$ is the angular frequency of the wave. Neglecting terms that are quadratic in the wave amplitude, one obtains:

\begin{equation} \label{eq:model1alinear}
 (\sigma-\mathrm{i}\omega)A=\tilde X A \left(\exp(\mathrm{i}\kappa)-\exp(-2\mathrm{i}\kappa) \right)-\alpha B -\gamma A 
\end{equation}
\begin{equation} \label{eq:model1blinear}
  (\sigma-\mathrm{i}\omega)B=2\mathrm{i}\tilde X B \sin(2\kappa)+2\mathrm{i}\tilde \theta A \sin(\kappa)+  \alpha A-\gamma B,
\end{equation}
We exclude the trivial solution $A=B=0$ and, thanks to linearity, we set $A=1$ (only the {\color{black}ratio} $b=B/A$ is indeed relevant). We separate real and imaginary part in the previous equations and obtain: 
\begin{equation} \label{eq:model1alinear1}
 \sigma=\tilde X \left(\cos(\kappa)-\cos(2\kappa)\right) -\alpha \mathbf{Re} \{b\} -\gamma 
\end{equation}
\begin{equation} \label{eq:model1alinear2}
 \omega=-\tilde X  \left(\sin(\kappa)+\sin(2\kappa)\right) +\alpha \mathbf{Im}\{b\} 
\end{equation}

\begin{equation} \label{eq:model1blinear1}
  \sigma \mathbf{Re}\{b\}+\omega \mathbf{Im}\{b\} =-2\tilde X \mathbf{Im}\{b\} \sin(2\kappa)+ \alpha -\gamma \mathbf{Re}\{b\},
\end{equation}

\begin{equation} \label{eq:model1blinear2}
   \sigma \mathbf{Im}\{b\}-\omega \mathbf{Re}\{b\} =2\tilde X \mathbf{Re}\{b\} \sin(2\kappa)+2\tilde \theta  \sin(\kappa)-\gamma \mathbf{Im}\{b\},
\end{equation}
where $\mathbf{Re}\{x\}$ and $\mathbf{Im}\{x\}$ are the real and imaginary part of the complex number $x$, respectively. {\color{black}The conditions leading to the bifurcation associated with the loss of stability of the fixed point given in Eqs. \ref{fixedpointa}-\ref{fixedpointb} can be derived by setting $\sigma=0$ in Eqs. \ref{eq:model1alinear1}-\ref{eq:model1blinear2} and finding $\omega$, $\kappa$, $\mathbf{Re}\{b\}$ and $\mathbf{Im}\{b\}$ and a function of the parameters $F$, $G$, $\alpha$, and $\gamma$. In the case $\omega\neq0$, the onset of the neutral wave  corresponds to  a Hopf bifurcation. }

Solving the previous Eqs. \ref{eq:model1alinear1}-\ref{eq:model1blinear2} and finding the expression of $\sigma$ and $\omega$ as a function of $\kappa$ and of the parameters $F,G,\alpha$, and $\gamma$ gives the dispersion relation of the waves. Additionally, obtaining the real and imaginary part of $b$ allows for understanding the relative amplitude of the waves in the $X$ and $\theta$ variables. 

{\color{black}Note that the linear stability analysis of the L96 model can be obtained by setting  $\alpha=0$, $\gamma=1$ in Eq. \ref{eq:model1alinear} and neglecting, instead, the $\theta$ variables. One then recovers the result first presented in \cite{Lorenz2005} and discussed in greater detail in \cite{vanKekem2018PhysD}. {\color{black}In what follows we consider $F\geq0$; an analysis of the somewhat dynamics occurring for $F<0$ has been presented in \cite{vanKekem2019}.} It is possible to derive the minimal value of $F$ such that the fixed point of the system loses stability and, correspondingly, to obtain the wavelength and frequency of the emerging neutral wave.} One finds that, taking a continuum approximation ($N\rightarrow\infty$){\color{black}, the neutral wave is realised when  $F=F_{crit}=8/9$, where the critical wavenumber is $\kappa=\kappa_{crit}=\arccos(1/4)$, and the critical frequency is $\omega_{crit}=-F_{crit}(\sin(\kappa_{crit})+\sin(2\kappa_{crit}))\approx -1.29$.} {\color{black} If one assumes that the gridpoints {\color{black}are arranged like along a latitudinal circle where the longitude increases with the index of the gridpoints (note that the periodic boundary conditions of the system impose a toroidal topology)}, we have that the crest of the neutral wave moves westward, because the phase velocity $v_p=\omega_{crit}/\kappa_{crit}= \approx-0.98$ is negative. Instead, the group velocity $v_{g}=\partial \omega_{crit}/\partial \kappa|_{\kappa=\kappa_{crit}}=-F_{crit}(\cos(\kappa_{crit})+2\cos(2\kappa_{crit})) \approx 1.33>0$ 
so that the wavepackets have an eastward propagation.}

As a result of the presence of the coupling between the $X$ and $\theta$ variables, it is hard to find {\color{black}for the model introduced in this paper} an explicit expression for the conditions supporting the presence of a neutral wave, also if one takes the special cases where one between $F$ and $G$ vanishes. A simple solution is instead found if one takes $\gamma=\alpha=1$ and $F=G$, which implies $\tilde X = 0$ and $\tilde \theta = F$. One then obtains the following results when imposing $\sigma=0$ {\color{black} and taking the continuum approximation}: $\mathbf{Re}\{b\}=-1$, $\omega= \mathbf{Im}\{b\}=\sqrt{2}$, $\kappa=\arcsin(\sqrt{2}/F)$. This indicates that $F_{crit}=\sqrt{2}$, and $\kappa_{crit}=\pi/2$ (corresponding to a critical wavelenght of 4), and $\omega_{crit}=\sqrt{2}$. {\color{black}Therefore, the phase velocity of the neutral wave $v_{p}=\omega_{crit}/\kappa_{crit}=2\sqrt{2}/\pi$ is positive, corresponding to an eastward motion of the wave crests. Since $\omega_{crit}=F_{crit}\sin(\kappa_{crit})$, we have $v_{g}=\partial \omega_{crit}/\partial \kappa|_{\kappa=\kappa_{crit}}=F_{crit}(\cos(\kappa_{crit}))=0$, implying no net motion of the wave packets.}

\section{Results}\label{Tests}

Many are the possible scientific questions one can address regarding the model introduced above. Building on the large literature on the L96 model discussed in the introduction, and taking into account the extra features of the current model, we can mention the following lines of investigation:
\begin{itemize}
    
     \item Analysis of the bifurcations leading the system from fixed point to a periodic and quasi-periodic behaviour to a chaotic regime as the forcing is increased;
      \item Systematic investigation of the predictability of the system - e.g. analysis of the finite-time and asymptotic Lyapunov exponents and the corresponding covariant Lyapunov vectors as a function of the two forcing parameters $F$ and $G$;
    \item Systematic investigation of the energetics of the system as a function of the two forcing parameters $F$ and $G$;
     \item Analysis of the signal propagation through waves in the quasi-periodic and weakly chaotic regime;
    \item Definition of asymptotic scaling laws for the properties if the system for large values of $F$ and $G$;
    \item Detection and analysis of chaos extensivity as the number of gridpoints $N\rightarrow \infty$;
    \item Extension of the model to multiple scales and analysis of dynamics and of the energetics of scale-to-scale interaction.
\end{itemize}
Obviously, it is impossible to address with a high level of detail all these aspects in the present paper. Rather than focusing on one or few aspects among those above, since this is the first time this model is proposed to the scientific community, we  will present some preliminary results that address partially each of the points above, in the hope of stimulating a reader into going in greater detail. Further studies by the authors that focus specifically in some of the aspects mentioned above will be reported elsewhere.

All the numerical integrations are performed using a Dormand–Prince method with adaptive time step and a spin time of $100$ time units, with runs of $1000$ time units. We make use of the Python module JiTCODE \cite{Ansmann2018}, an extension of SciPy’s ODE that allows to numerically simulate ordinary differential equations, computing quantities of interest as Lyapunov exponents as well. All results have been double checked and confirmed using the MATLAB function \texttt{ode45} where integrations are performed using the 4$^{th}$ order Runge-Kutta integrator with adaptive time step \cite{Shampine1997}.

\subsection{Transition to Chaos and Predictability}

\begin{figure}
a)\includegraphics[width=.48\linewidth]{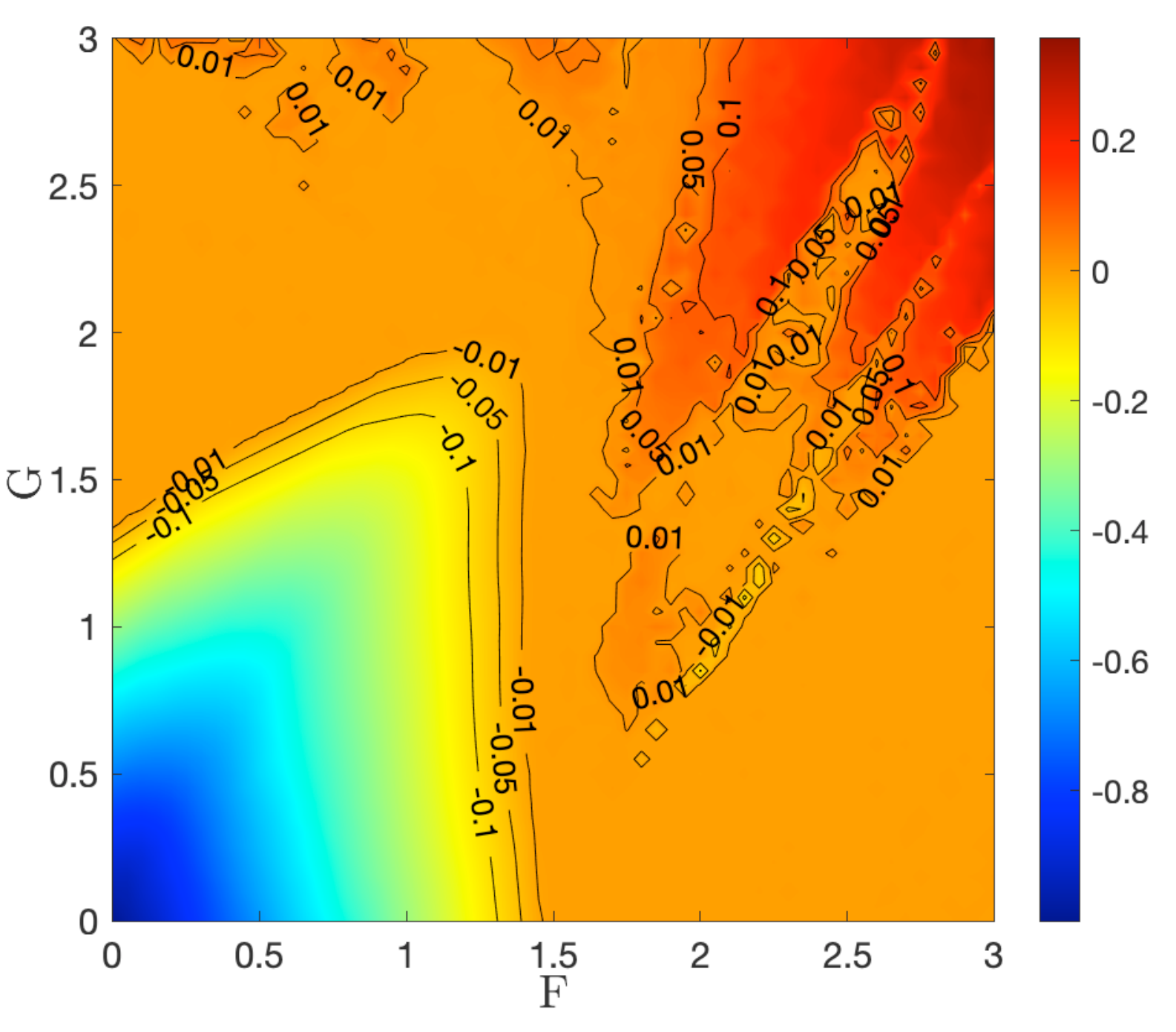}
b)\includegraphics[width=.48\linewidth]{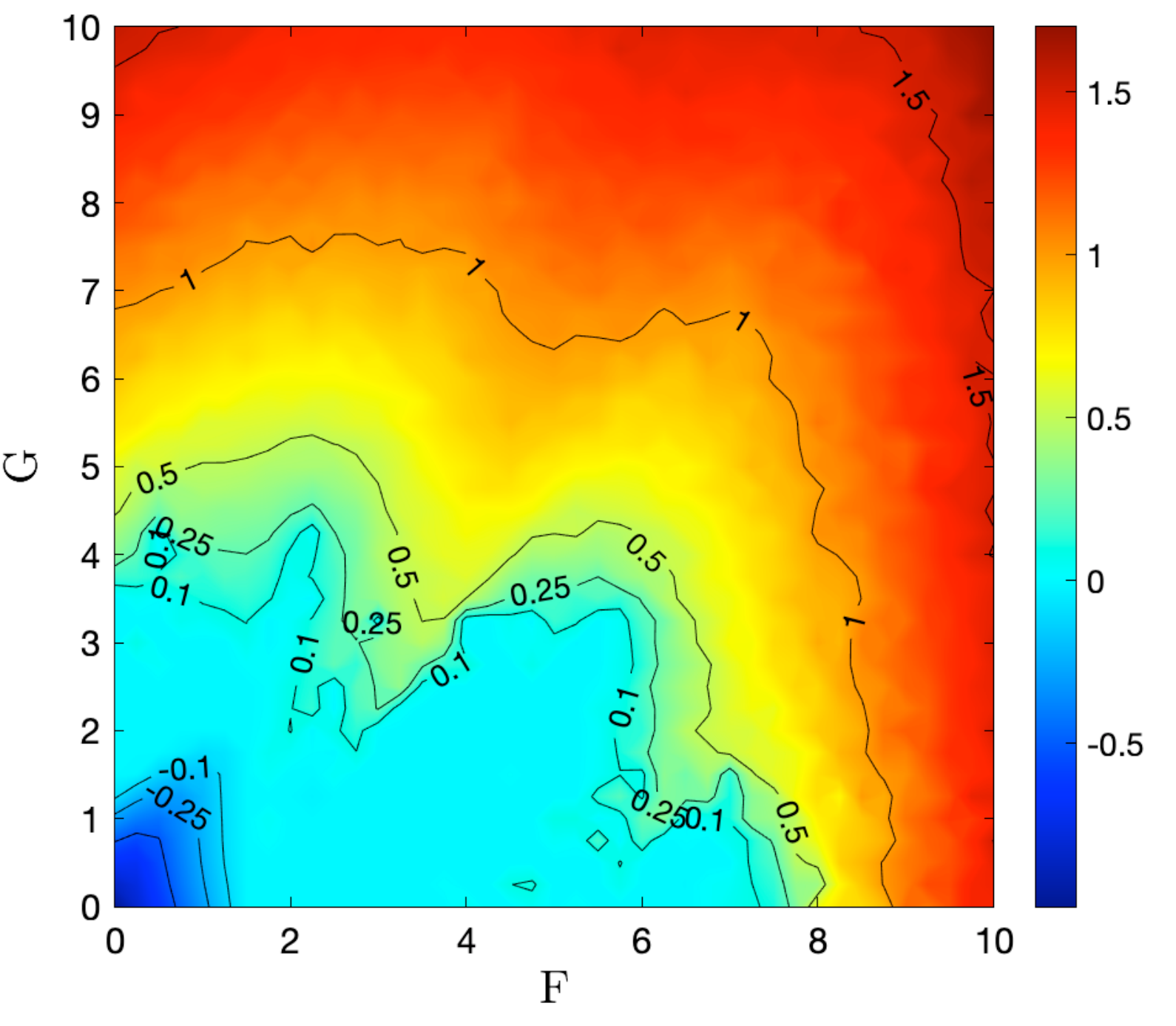}
\caption{\label{Figure_3} First Lyapunov exponent of the one-level model with $N=36$ gridpoints as a function of  $F$ and $G$. a) Detail for $0\leq F,G\leq3$. b) Full range for $0\leq F,G\leq10$. {In all simulations \color{black}$\gamma=\alpha=1$}.}
\end{figure}

A simple yet fundamentally correct way to characterise at qualitative level the dynamical properties of a system is to investigate to what extent its evolution is sensitive to its initial conditions. Roughly speaking, the first Lyapunov exponent of a system measures the asymptotic rate of growth or decay of the distance between two orbits which are initialised in the attractor of the system at infinitesimal distance from each other \cite{Strogatz2014}. Similarly, one can define the sum of the first $p$ Lyapunov exponents as defining the asymptotic average rate of growth or decay of the $p$-volume defined by $p+1$ orbits that are initialised in the attractor of the system at infinitesimal distance from each other \cite{Eckmann1985}. Indeed, for a $Q$ dimensional continuous time dynamical system, it is possible to compute $Q$ Lyapunov exponents $\lambda_1,\ldots,\lambda_Q$, where the customary ordering is such that $\lambda_j\leq\lambda_k$ if $k\leq j$ of the  \cite{Benettin1980}. 

If $\lambda_1<0$ the attractor is a fixed point, whilst if $\lambda_1=0$ the attractor is periodic or quasi-periodic. Finally, the presence of a positive first Lyapunov exponent is a  significant evidence that the system is chaotic, and the value of such exponent determines quantitatively the rapidity with which two nearby trajectories diverge from each other. In this case, one has that there is at least one $\lambda_j=0$, $j>1$, which corresponds to the direction of the flow  \cite{Eckmann1985,Strogatz2014}. 

Figure \ref{Figure_3} shows the estimate of $\lambda_1$ for $N=36$ as a function of $F$ and $G$ in the range $0\leq F, G \leq 10$. We remark that the Python JiTCODE module allows for the computation of the full spectrum of Lyapunov exponents using the algorithm proposed in \cite{Benettin1980}. The system has a negative  $\lambda_1$ for small values of the forcings, as expected, see Fig. \ref{Figure_3}a). We remark that if $G=0$ $\lambda_1\leq 0$ for $F\leq 1.4$, whereas for the L96 model $F_{crit}=8/9$, indicating that presence of a mechanism of energy transfer between kinetic and potential energy and the presence of a new channel of dissipation (for potential energy) leads to higher stability for the system. We observe that  $\lambda_1$ depends in a very nontrivial way on both $F$ and $G$, as the system's behaviour depends delicately on how the energy is injected into it, because the dynamics of the $X$ and $\theta$ variables is, in fact, quite distinct. It is extremely different to force the system through kinetic or the potential energy channel. We also observe that the theoretical prediction of $F_{crit}=\sqrt{2}$ for $F=G$ agrees with what shown in Fig. \ref{Figure_3}a). 

Many other interesting features appear. Increasing $F$ from zero to 3 while keeping $G=1.7$, the asymptotic dynamics of the system changes first from quasi-periodic to a fixed point, then again to quasi-periodic, which then alternates with chaotic behaviour. Indeed, one can observe two complex tongue-like structures in Fig. \ref{Figure_3}a) for $F\geq 1.7$, $G\geq 0.5$, which indicate the presence of a very nontrivial set of bifurcations for that regions of the parameters' space, defining the transition between the quasi-periodic behaviour - the light orange region - ad the chaotic regime - the dark orange and red region in Fig. \ref{Figure_3}a).

Zooming out towards a larger range of values for $F$ and $G$ the  intuitive argument that increasing either $F$ or $G$ makes the system less predictable becomes more robust, even though there are regions where a destructive interference is clear (in terms of values of $\lambda_1$ between the two forms of forcing, compare the two troughs near the diagonal in  Fig. \ref{Figure_3}b). 

{\color{black}We remark that it is reasonable to expect that, as in the case of the L96 model \cite{vanKekem2018NPG,vanKekem2018PhysD,vanKekem2019}, in the regime of moderate forcing the position and nature of the bifurcations will depend delicately on the number of nodes $N$, so that one should expect modifications especially in  Fig. \ref{Figure_3}a) when performing simulations for a value of $N$ other than 36 considered here. Instead, as shown below in Sect. \ref{extensivity}, one finds some indication of universality associated with the continuum limit $N\rightarrow\infty$ when sufficiently strong forcing is considered.}

\subsection{Energetics}

\begin{figure}
a)\includegraphics[width=.45\linewidth]{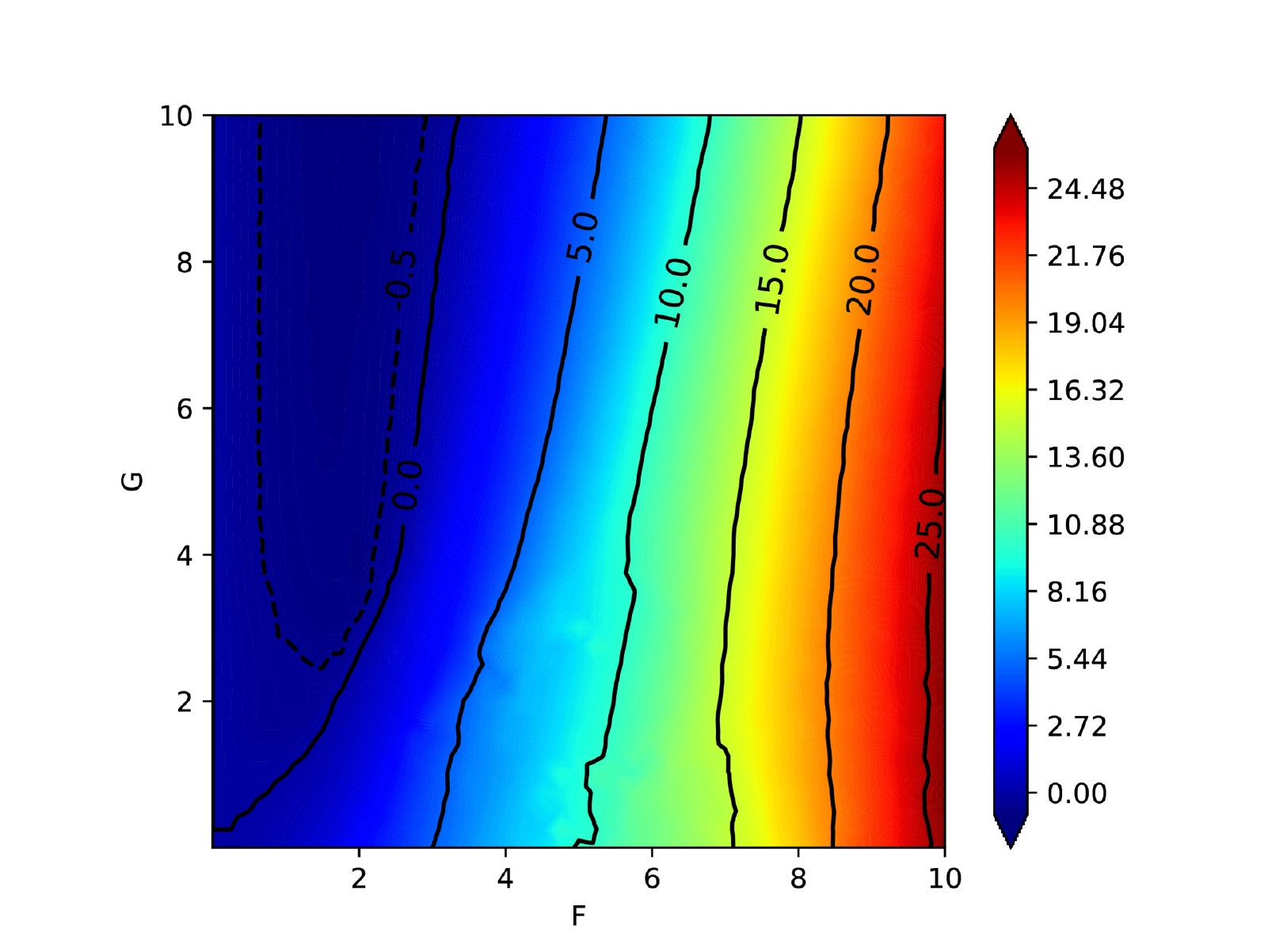}
b)\includegraphics[width=.45\linewidth]{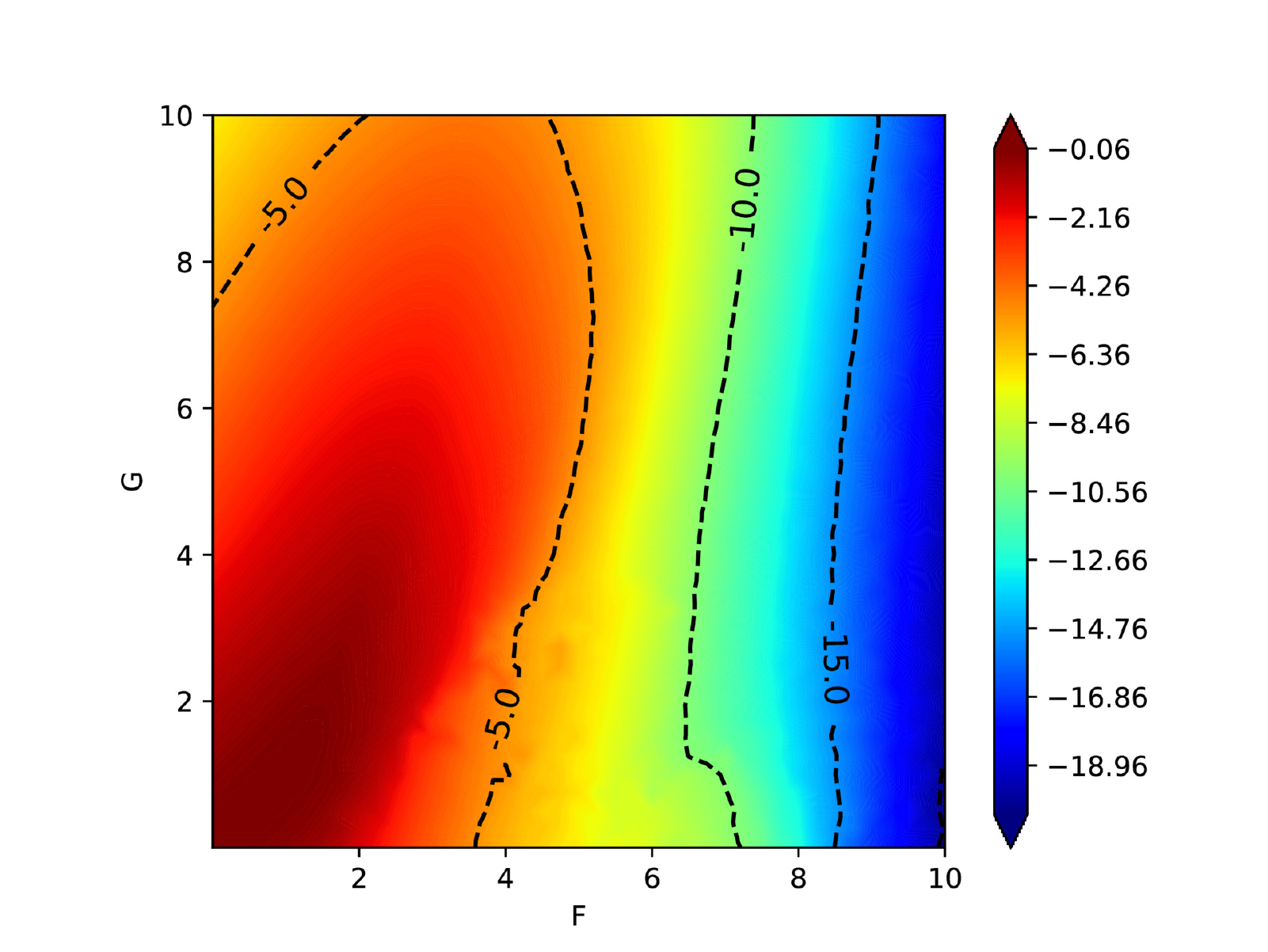}\\
c)\includegraphics[width=.45\linewidth]{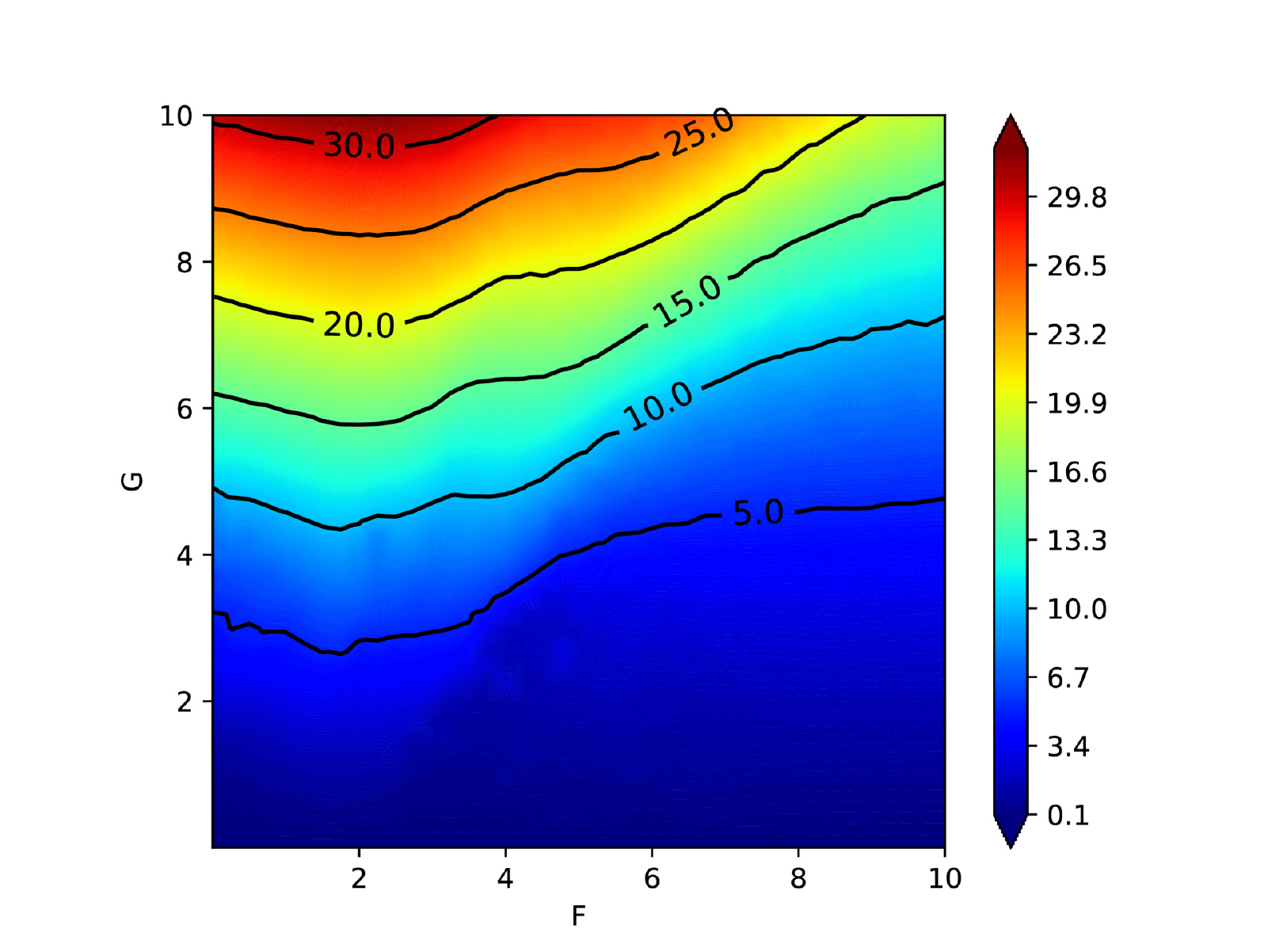}
d)\includegraphics[width=.45\linewidth]{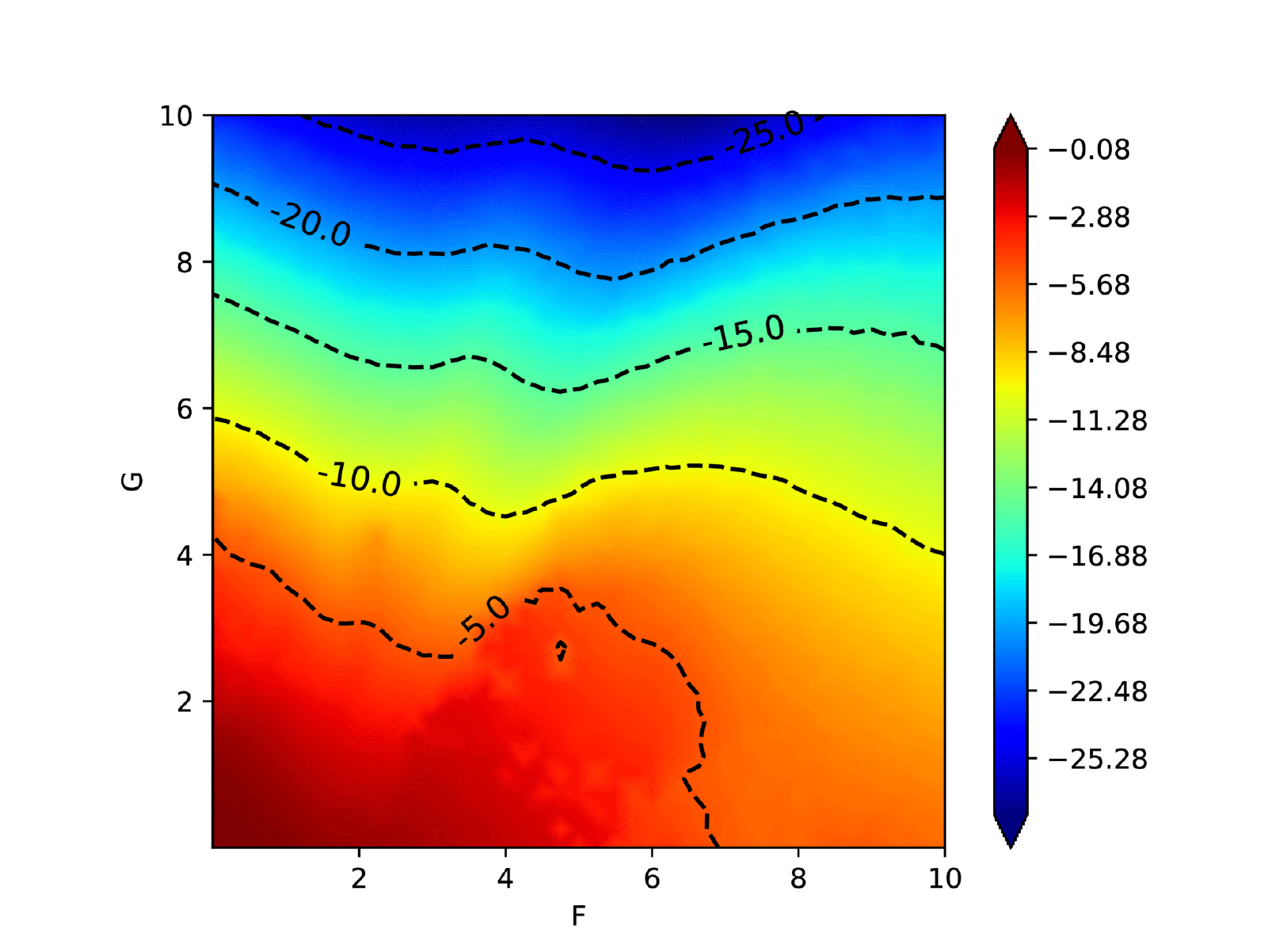}\\
e) \includegraphics[width=.45\linewidth]{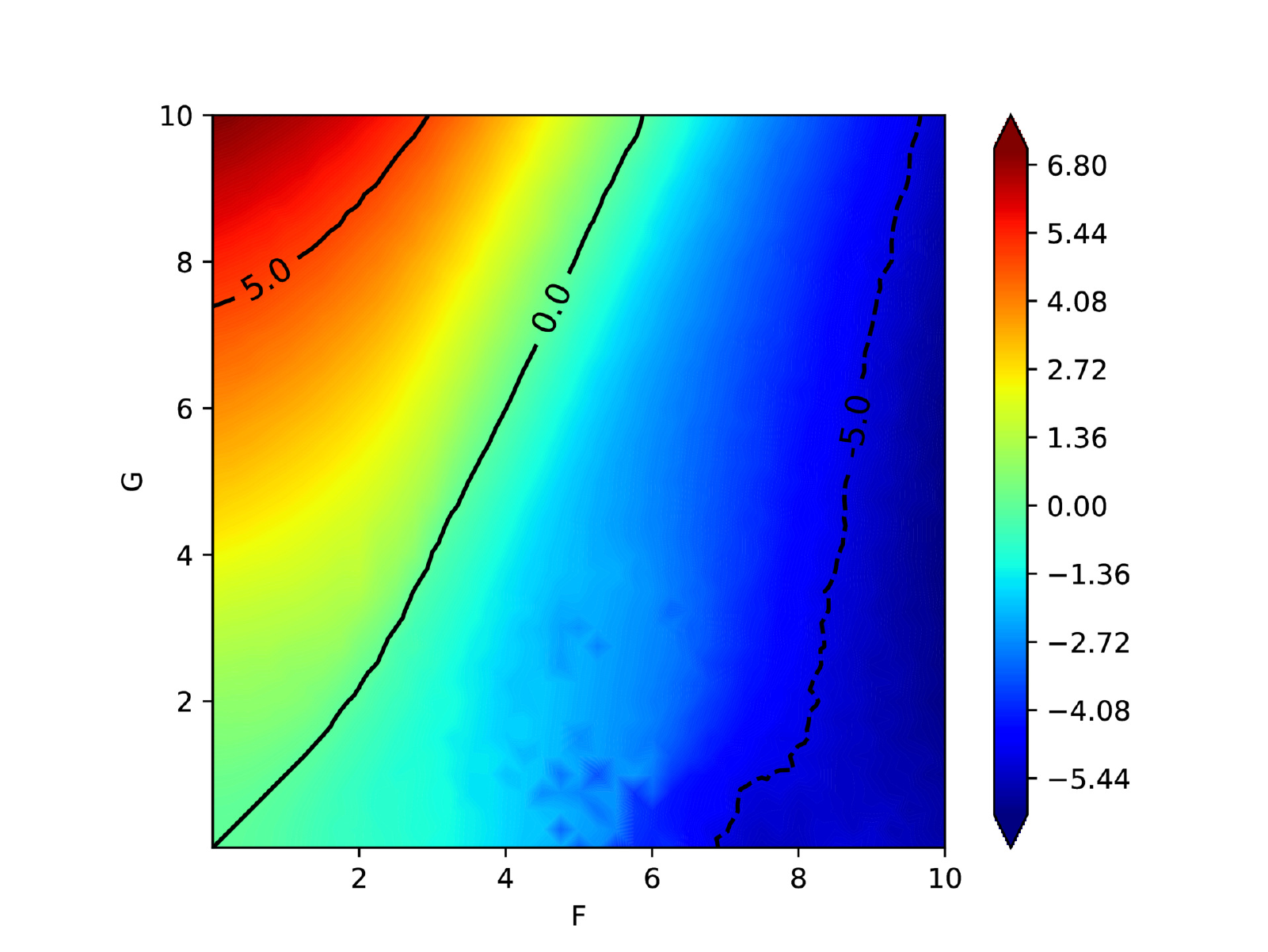}
f) \includegraphics[width=.45\linewidth]{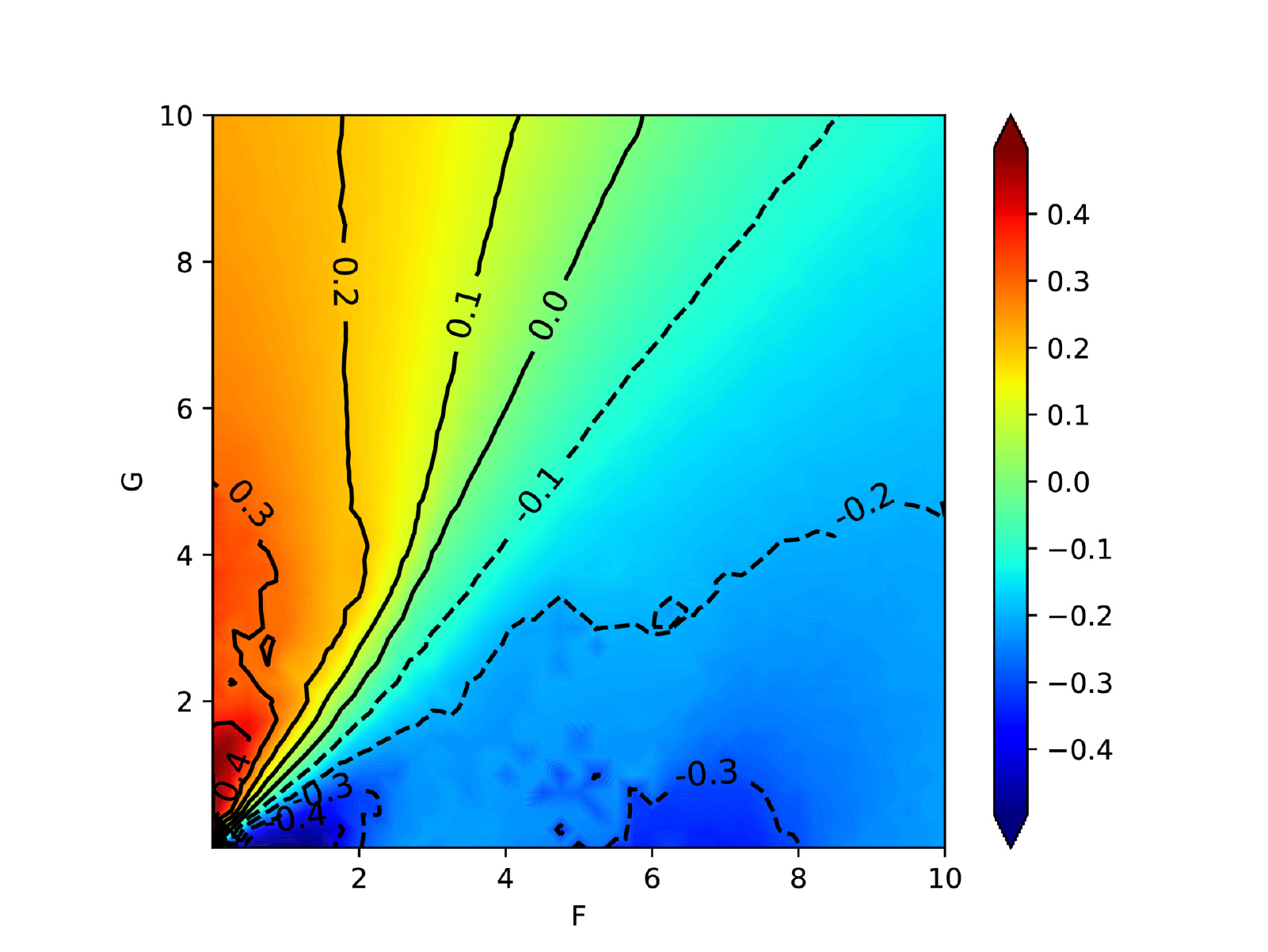}
\caption{\label{Figure_31} a) $F \frac{1}{N} \bar \Xi = F \frac{1}{N}\sum_{k=1}^N X_k$; b) $-2\gamma \frac{1}{N} \bar K = -\gamma \frac{1}{N}\sum_{k=1}^N X_k^2$; c) $G \frac{1}{N} \bar \Theta = G \frac{1}{N}\sum_{k=1}^N \theta_k$; d) $-2 \frac{1}{N} \gamma \bar P = -\gamma \frac{1}{N}\sum_{k=1}^N \theta_k^2$; e) $\frac{1}{N} \bar C=-\frac{1}{N}\sum\limits_{k=1}^N \left(\alpha X_k \theta_k \right)$; f) $\eta=\bar{C}/(2\gamma\bar{E})$ as a function of $F$ and $G$. {\color{black}In all simulations $\gamma=\alpha=1$}.}
\end{figure}

It is useful to investigate the long-term average of the terms in Eqs.\ref{LEC1}-\ref{LEC2} as a function of $F$ and $G$, see  Fig.\ref{Figure_31}. The lack of equivalence between applying forcing to the $X$ vs to the $\theta$ variables is extremely clear by looking at the conversion term (panel e). $\bar C$ is positive for large values of $G$ and moderate values of $F$, and negative viceversa. The absolute value of $\bar C$ increases with $F$ ($G$) if $G$ ($F$) is kept constant. The zero isoline strongly deviates from the diagonal and indicate that if $F=G$ there is a net transfer of energy from kinetic to potential. The zero isoline of $\bar C$  coincides with the ridge in the value of $\lambda_1$ shown in \ref{Figure_3}b), indicating  that the condition of no net energy exchange between the two reservoirs of energy corresponds to a state where instabilities are rather strong. The zero-isoline of the efficiency $\eta$ (see panel f), by definition, coincides with the one of $\bar{C}$. The absolute value of the efficiency grows with the asymmetry of the forcing, and peaks for moderate intensity of either $F$ or $G$, suggesting - see Sect. \ref{scaling} below - that the energy conversion becomes less efficient  decreases when stronger forcings are considered.

The behaviour of the other thermodynamical quantities is somehow unsurprising, as we have that both input and dissipation of kinetic (potential) energy increase with $F$ ($G$). We remark that, once again, the response of the system to the two individual forcings is quantitatively different. It should be noted that, when one considers $F\leq2$, for $G\geq2$ one has that the net input of kinetic energy is negative (with the dissipation of kinetic energy, being, by definition, positive). This indicates a very nontrivial impact of the thermodynamic variables on the dynamical ones, which are the only ones performing advection. As a result, there is an additional mechanism of energy loss for the system, whilst all the energy input takes place through the potential energy channel. Instead, when considering low values of $G$, the potential energy input is always positive - yet small. 


\subsection{Waves amidst Chaos}
\begin{figure}
a)\includegraphics[width=0.45\linewidth]{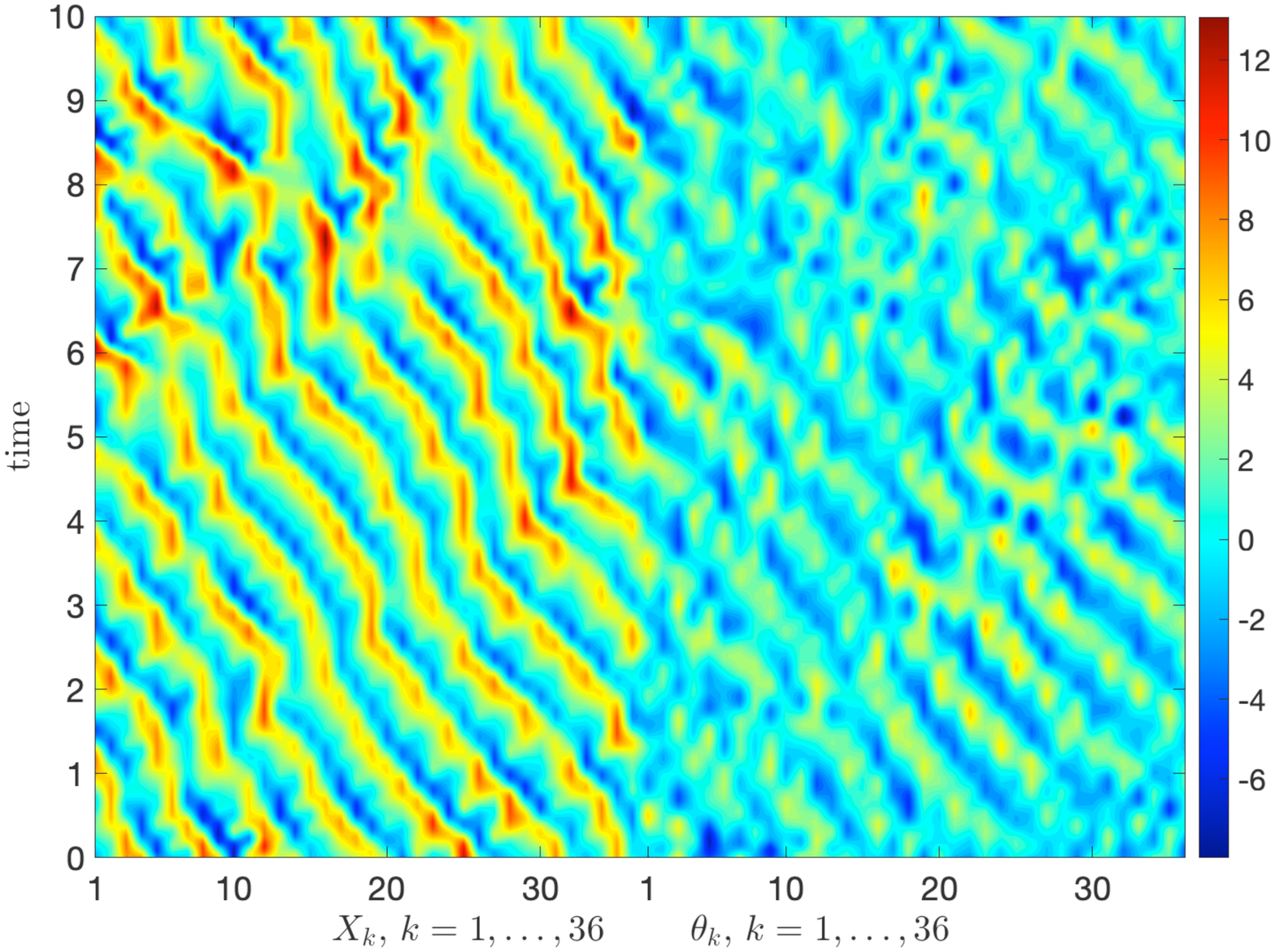}
b)\includegraphics[width=0.45\linewidth]{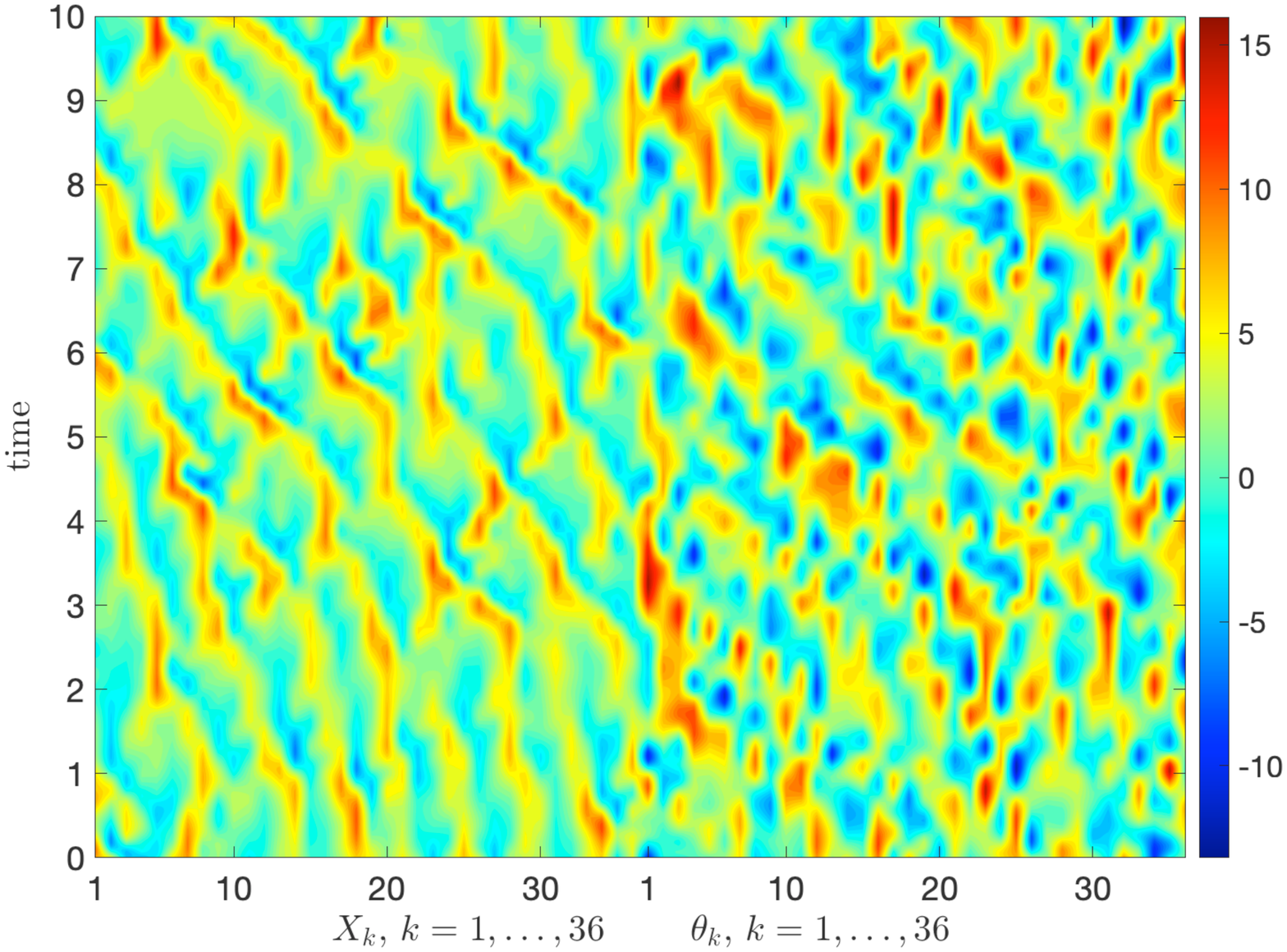}\\
c)\includegraphics[width=0.45\linewidth]{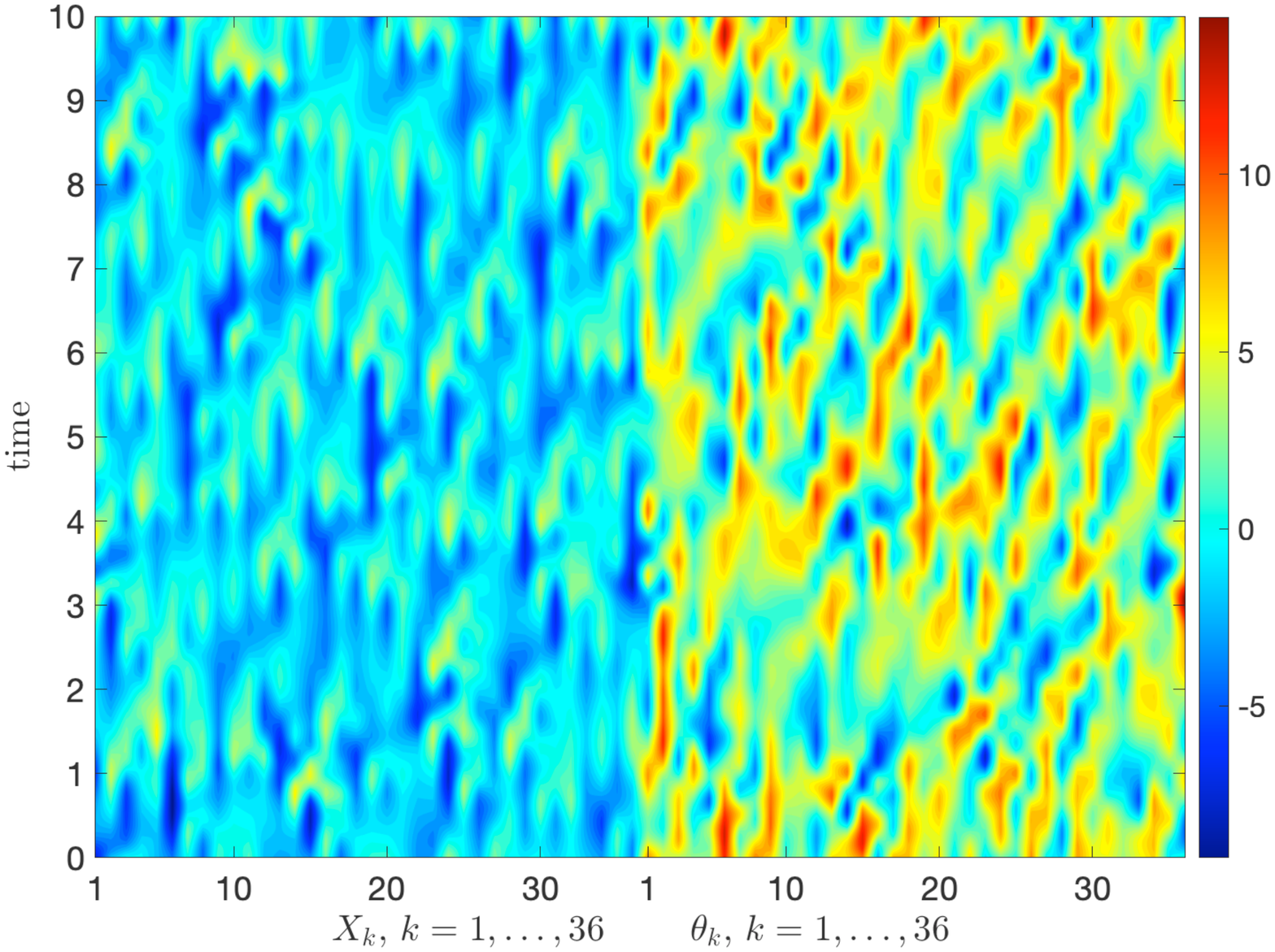}
d)\includegraphics[width=0.45\linewidth]{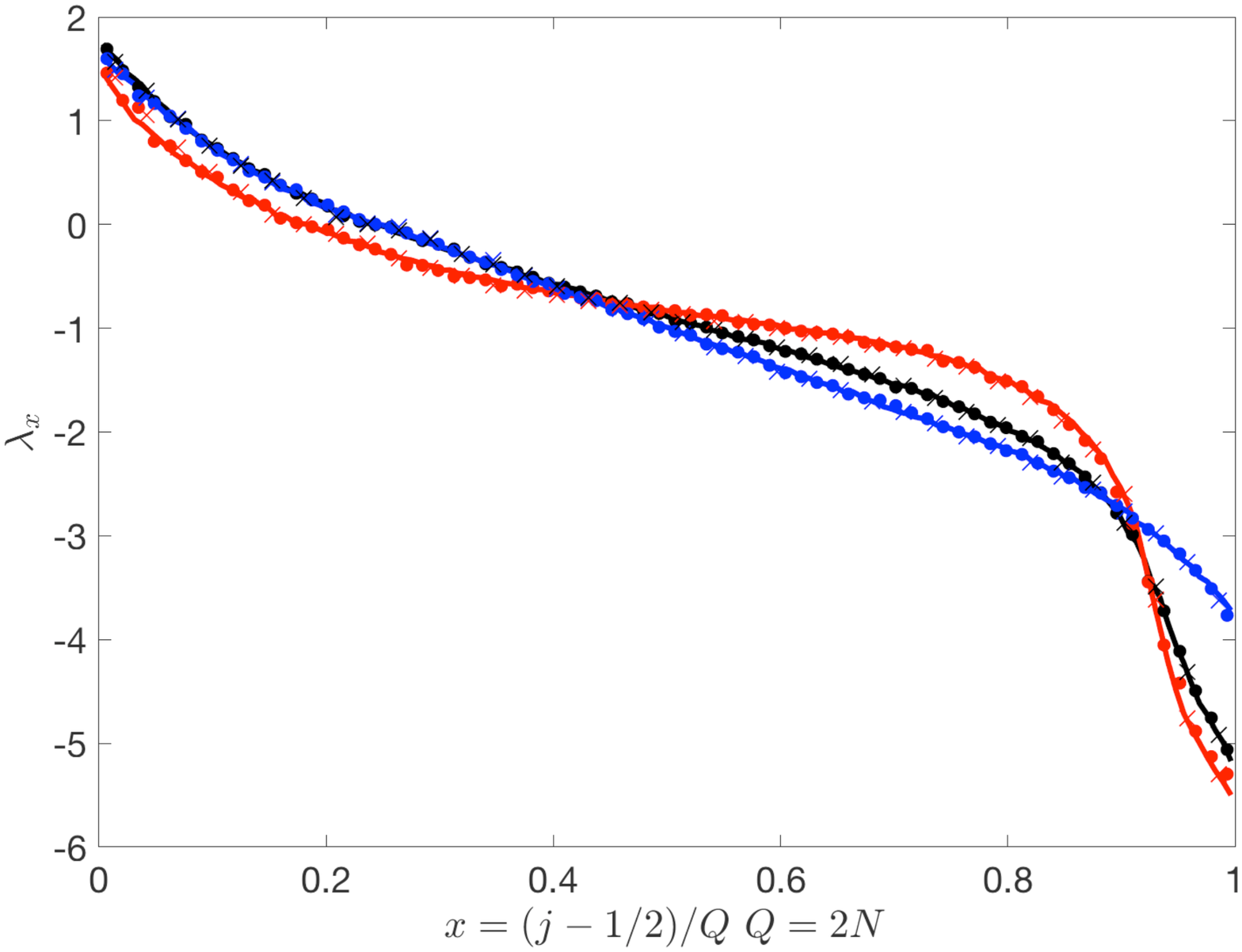}
\caption{\label{Figure_1} {\color{black}Slice of 10 time units of} the evolution of the system with $N=36$ for $F=10$, $G=0$ (panel a); $F=10$, $G=10$ (panel b); and $F=0$, $G=10$ (panel c). {\color{black}Time is on the vertical axis. On the horizontal axis, the first 36 variables are the $X_k$, $k=1,\ldots,36$, followed by the variables $\theta_k$, $k=1,\ldots,36$.} Panel d): Extensivity of the system - spectrum of Lyapunov exponents as a function of the rescaled index $x=(j-1/2)/(2N)$ for $F=10$, $G=0$  (red); $F=10$, $G=10$  (black); $F=0$, $G=10$  (blue). Solid lines: $N=72$; Dots: $N=36$; Crosses: $N=18$. In all simulations, $\alpha=\gamma=1$. }
\end{figure}

We highlight some qualitative features of the dynamics of the model that indicate the presence of wave-like structures amidst chaos in the regime of moderate forcing. Figure \ref{Figure_1} shows some examples of evolution of the system of the system in the case of $N=36$ sectors with $F=10$, $G=0$ (panel a); $F=10$, $G=10$ (panel b); and $F=0$, $G=10$ (panel c). {\color{black}We are using a Hovm\"oller-type diagramme \cite{hovmoller1949}, where time is on the vertical axis and the variables $X_k$ and $\theta_k$, $k=1,\ldots,36$ are on the horizontal axis. This diagramme is particularly well suited for appreciating wave-like structures, as it is easy to to visualise wave crests.} If the forcing on the $\theta$ variables is switched off, the $X$ variables behave similarly to the case of the L96 model, where, amidst chaos, the clear signature of a westward propagating phase velocity can be found, {\color{black}as already observed in \cite{Lorenz2005} and recently mentioned in \cite{Kerin2020}.}
As can be guessed from the evolution equations, the $\theta$ variables feature weaker variability and similar pattern of the wave crests, as they are advected by the $X$ variables and receive energy from them. The situations is qualitatively similar when both the $X$ and $\theta$ variables are forced, but, quite naturally, the fluctuations of the $\theta$ variables are stronger than in the previous case. {\color{black}Note that in the case analysed here of $F=G=10$, the wave crests travel in the opposite direction with respect to what we have found for the neutral wave emerging for $F=G=\sqrt{2}$, see Sect. \ref{linearstability}. Therefore, the presence of a turbulent background radically changes the kinematics of the waves}. If, instead, the forcing acts on the $\theta$ variables only, the wave crests have a much less clear direction of propagation, both for the $\theta$ and for the $X$ variables, where the latter feature a much lower variability, as expected.  {\color{black}In other terms,  the setup where $F$ vanishes is characterised by absolute instability, with little or no advection of anomalies, whereas the other two cases  are characterised by convective instability, where anomalies are spatially advected \cite{Huerre1990}.}

We will further discuss in the following sections in more quantitative terms the differences emerging when forcing the $X$ variables only, the $\theta$ variables only, or all variables.

\subsection{Chaos Extensivity} \label{extensivity}
Ruelle \cite{Ruelle1982} proposed that systems with short-range interaction can feature extensive chaos, because large domains can be hierarchically  partitioned into smaller, weakly interacting subdomains with similar properties. One way to test whether chaos extensivity is to analyse the finite-size scaling of the Lyapunov exponents. Specifically, one plots the obtained spectrum of Lyapunov exponents for different values of system size $Q$ (in our case, $Q=2N$) against the rescaled index $x=(j-1/2)/Q$  and tests whether a universal curve is obtained in the limit of large values of $Q$  \cite{Takeuchi2011,Gallavotti2014}.  We remark that chaos extensivity implies that the ratio between the  Kaplan-Yorke dimension of the attractor \cite{Kaplan1979}, also referred to as Lyapunov dimension \cite{Ott1993}, and $Q$ tends to a constant as $Q\rightarrow\infty$. 

In order to prove convincingly the extensive nature of chaos in the  system analyzed here,  one should test such property for all values of $F$ and $G$. This is beyond the scope of this paper. Yet, preliminary results do confirm extensivity for the three reference cases  $F=10$, $G=0$; $F=G=10$; $F=0$, $G=10$ shown in Panels a)-c) of Fig. \ref{Figure_1}. {\color{black}Indeed, we have here performed simulations with $N=18$, $N=36$, and $N=72$ and, as shown in \ref{Figure_1}d), the Lyapunov exponents spectra seem to collapse to universal curves as $N$ grows. Indeed, the Lyapunov spectra plotted against their respective rescaled indices can hardly be visually  distinguished.} This is especially encouraging in view of the clear evidence for chaos extensivity in the L96 model \cite{Karimi2010,Gallavotti2014}.

\subsection{Scaling Laws for Strong Forcings}\label{scaling}
As thoroughly analyzed in \cite{Gallavotti2014}, in the one-layer L96 model the average energy per unit site scales to a very high degree of approximation as $F^{1.33}$ for large values of $F$. The origin of such a scaling law is still unknown. We report some preliminary results of scaling laws obtained for the current model in some special configurations of parameters. We have performed long integrations (1000 time units) at steady state for three set of experiments:
\begin{enumerate}
    \item $F=2^j$, $j=3,\ldots,14$; $G=0$
    \item $F=G=2^j$, $j=3,\ldots,14$;
     \item $G=2^j$, $j=3,\ldots,14$; $F=0$
\end{enumerate}
which correspond to applying a forcing of increasing strength on the $X$ variables only, on both the $X$ and the $\theta$ variables, or on the $\theta$ variables only, respectively. {\color{black}These are regimes of forcing where, see the case of the L96 model \cite{Gallavotti2014}, one might expect that chaos extensivity applies with a very good approximation, see Sect. {\color{black}\ref{extensivity}}.}  We obtain the following approximate asymptotic scaling laws, which are rather accurate when $F$ and/or $G$ are larger than 256: 
\begin{enumerate}
    \item $\bar{K},\bar{E}\propto F^{1.33}$, $\bar{\Xi}\propto F^{0.33}$, $\bar{\Theta}\propto F^{-0.28}$,  $\bar{P}=-\bar{C}/2\propto F^{0.71}$, $\lambda_1\propto F^{0.66}$;
    \item $\bar{K},\bar{E},\bar{P}\propto F^{1.33}$, $\bar{P}\approx 0.7\bar{K}$, $\bar{C}<0,|\bar{C}|\propto F^{0.70}$, $\bar{\Xi},\bar{\Theta}\propto F^{0.33}$, $\bar{\Theta}\approx 0.7\bar{\Xi}$, $\lambda_1\propto F^{0.66}$;
     \item $\bar{P},\bar{E}\propto G^{1.50}$, $\bar{K}=\bar{C}/2\propto G^{1.00}$, $\bar{\Theta}\propto G^{0.50}$, $\bar{\Xi}\propto G^{0.00}$, $\lambda_1\propto G^{0.50}$;
\end{enumerate}
where the uncertainty is 0.01 for all the numbers above. As clear from these scaling relations, and in agreement with what one could intuitively guess by looking at Fig. \ref{Figure_1}, it is rather different to force the system through the $X$ or the $\theta$ variables, and the interplay between the two reservoirs of energy is far from trivial. If the forcing is applied to only one set of variables, the energy cycle is more enhanced, \textit{ceteris paribus}, when the $\theta$ variables undergo the forcing. Indeed, the reservoir of total energy and the conversion $\bar{C}$ of energy between the two forms are larger than for corresponding case of forcing applied uniquely to the $X$ variables. The behaviour of the quantities $\bar{\Xi}$ and $\bar{\Theta}$ is also extremely different in the two cases, implying a qualitatively different way the forcing impacts the spatially-coherent fluctuations of the variables.  If $F=G$, the ratio of the average size of the two reservoirs of energy is a constant, with $\bar{K}$ being larger that $\bar{P}$ (and, correspondingly, $\bar{\Xi}$ being larger than $\bar{\Theta}$), and the average flux of energy goes from kinetic to potential. {\color{black}We remind that the dynamics of the $F=G$ case is characterised by convective instability, similarly to the case where $G=0$, compare \ref{Figure_31}a) and \ref{Figure_31}b). The reason why the forcing through the kinetic channel dominates in the special case of $F=G$ is still unclear and should be further investigated.}

In all cases, the amount of energy that is converted between the two forms becomes a negligible fraction of the total incoming energy in the limit of large forcing. In other terms, the efficiency of the model $\eta=\bar{C}/{(2\gamma \bar{E})}$ tends to zero if either $F$ or $G$ tend to infinity, even if $\bar{C}$ tends to infinity. It is then unsurprising that when considering the limit of large $F$, regardless of whether $G$ is also increased, we obtain for the $X$ variables results that are in agreement with what featured by the one-layer L96, compare with \cite{Gallavotti2014}. At this regard, a useful piece of information is obtained by looking at the properties of $\lambda_1$ in the large forcing limit. One obtains that in scenarios 1 and 2, $\lambda_1\propto F^{0.66}$, which is again in excellent agreement with what obtained for the L96 model (including  the pre-exponential factor). Scenarios 1) and 2) seem like featuring a rather  similar dynamics,  the main difference between the two being the strength of the fluctuations of the $\theta$ variables; compare with panels a) and b) of Fig. \ref{Figure_1}.

The growth of $\lambda_1$ with $G$ is slower in the case $F$ is set to zero, {\color{black}where we have absolute instability.} We can gain a qualitative understanding of the different impact on $\lambda_1$ of changes in the value of $G$ vs $F$ by comparing panels a), b) and c) of Fig. \ref{Figure_1}, which nonetheless describe weaker regimes of forcing (what follows stands also in the case of stronger applied forcing).  

\section{Conclusions} \label{Conclusions}

Simple and conceptual models have proved extremely useful for  better understanding the dynamics of climate as a whole as well as of its individual components. Indeed, their usefulness spans from being the testbeds for developing new methods in terms of data analysis, data assimulation, and model testing; to supporting the definition of new metrics for testing more complex models; to providing valuable insights in the basic active physical mechanisms and most prominent mathematical features.

The model presented in this paper goes in this direction and has been constructed in order to provide a new layer of physical complexity to the L96 model by adding a new variable to each gridpoint of the model. This variable can be loosely interpreted as a local temperature and allows for the establishment of a complex energetics for the system, encompassing energy input, output, and conversion. Two forms of energy are present in the system, a kinetic one and potential one.  We are also able to introduce a notion of efficiency, which is useful for studying the conversion of energy from one form to the other one. {\color{black}The energetics of the model is reminiscent of the one of the real atmosphere}. Extending previous analyses, we have provided a fairly complete analysis of the mechanics of the new model by separating a quasi-symplectic and a metric component to its dynamical structure. {\color{black}The energy of the system is used to construct the antisymmetric evolution operator, whose corresponding brackets are not true Poisson brackets because they do not obey the Jacobi identity, hence the symplectic structure is not complete.}

We have then performed a preliminary analysis of some of the key aspects of the new model by investigating how its properties change as a function of the two parameters that control the input of kinetic and potential energy.{\color{black} We have studied, in a special case, the Hopf bifurcation leading to the onset of the neutral wave from the fixed point solution.} The interplay between the two forcings is extremely non-trivial  in the weak forcing regime, where much needs to be explored regarding the transition from fixed point to quasi-periodic to chaotic asymptotic states, {\color{black} and one expects that the structure and position of the bifurcations might depend delicately on the number of modes included in the system, similarly to the case of the L96 model. When considering regimes associated with stronger forcing, the system exhibits extensive chaos, even if there is clear evidence of wave-like structures emerging in the context of an overall strongly chaotic flow. Understanding the interplay between ordered wave-like structures and turbulence seems of great interest.} 

The system reacts differently depending on how we force it. The nature of the flow is impacted because absolute vs convective instability dominate if we force the system through the potential energy vs kinetic energy channel, respectively. The mechanism of energy conversion makes sure that also the variables that are not directly forced feature nontrivial variability. If the strength of the forcing is the same in the two channels, the kinetic energy channel ends up being more efficient: the dynamics is characterised by convective instability, and, on the average, energy is transferred from the kinetic to the potential form. The reason for this behaviour is still unclear. Similarly to the case of the L96 model, it is possible to obtain  accurate power laws describing how some of the fundamental dynamical and thermodynamical properties of the system scale with the forcing parameters in the limit of very strong forcings. 

The analysis presented here is only a first step in the direction of better understanding the properties of this model, which  we believe has the potential of being of great interest for investigations in areas like  statistical physics, nonlinear dynamics, data assimilation, mechanics, model reduction techniques, and extreme events.



Finally, again along the lines of the L96 model, we have introduced a two-level version - see Appendix \ref{App_NM} - of the model, which allows for studying multi-scale dynamics and which features an energetics that resembles, conceptually, the one of the atmosphere, where the Lorenz energy cycle describes succinctly the input and output of energy in the kinetic and potential form as well as the conversion between the two forms and between energy compartments at small vs large scales. {\color{black}The study of the properties of this model, which is \textit{a fortiori} extremely promising in the fields  above, will be carried out in a future work.}


\appendix






\section{Two-level new model} \label{App_NM}
Here we propose  an extension of the model able to represent multiscale dynamics and energy exchanges across scales. A detailed investigation of the properties of this model will be discussed elsewhere. We present the evolution equation and discuss the energetics of the model. Mimicking the structure of the classical two-layer L96 model, we define the following evolution equations:
\begin{equation} \label{eq:model2a}
  \frac{dX_k}{dt}=X_{k-1}(X_{k+1}-X_{k-2})-\alpha\theta_k-\gamma X_k+F-\frac{hc}{b}\sum\limits_{j=1}^JY_{j,k},
\end{equation}
\begin{equation} \label{eq:model2b}
  \frac{dY_{j,k}}{dt}=cbY_{j+1,k}(Y_{j-1,k}-Y_{j+2,k})-\alpha c \phi_{j,k}-\gamma cY_{j,k}+f+\frac{hc}{b}X_k,
\end{equation}
\begin{equation} \label{eq:model2aa}
  \frac{d\theta_k}{dt}=X_{k+1}\theta_{k+2}-X_{k-1}\theta_{k-2}+\alpha X_k-\gamma \theta_k+G-\frac{hc}{b}\sum\limits_{j=1}^J\phi_{j,k},
\end{equation}
\begin{equation} \label{eq:model2bb}
  \frac{d\phi_{j,k}}{dt}=cb(Y_{j-1,k}\phi_{j-2,k}-Y_{j+1,k}\phi_{j+2,k})+\alpha c Y_{j,k}-\gamma c\phi_{j,k}+g+\frac{hc}{b}\theta_k,
\end{equation}
with $k=1,...,N$; $j=1,...,J$, where $Y_{j,k}$'s are $j$ small-scale variables coupled with $X_k$, similarly to the two-level L96 model, and $\phi_{j,k}$'s are $j$ small-scale variables similarly coupled with $\theta_k$. Additionally, the variables $Y_{j,k}$ and $\phi_{j,k}$ are also mutually coupled. The constant $hc/b$ determines the strength of the coupling between variables at different scale, while $c$ defines the time scale separation between the two levels and $b$ controls the relative amplitude of the fluctuations between large and small scales. Finally, $f$ ($g$) defines the forcing on the variables $Y_{j,k}$ ($\phi_{j,k}$). The boundary conditions are the following: 
\begin{equation}
\begin{split}
  X_{k-N}=X_{k+N}=X_k,\quad Y_{j,k-N}=Y_{j,k+N}=Y_{j,k},\quad  Y_{j-J,k}=Y_{j,k-1},\quad
  Y_{j+J,k}=Y_{j,k+1}.\\
  \theta_{k-N}=\theta_{k+N}=\theta_k,\quad
  \phi_{j,k-N}=\phi_{j,k+N}=\phi_{j,k},\quad
  \phi_{j-J,k}=\phi_{j,k-1},\quad
  \phi_{j+J,k}=\phi_{j,k+1}.
\end{split}
\end{equation}

Note that, choosing $\alpha=0$ and neglecting the $\theta$ and $\phi$ variables we obtain the classic two-layer L96 model.

\begin{figure}
\includegraphics[width=0.6\linewidth]{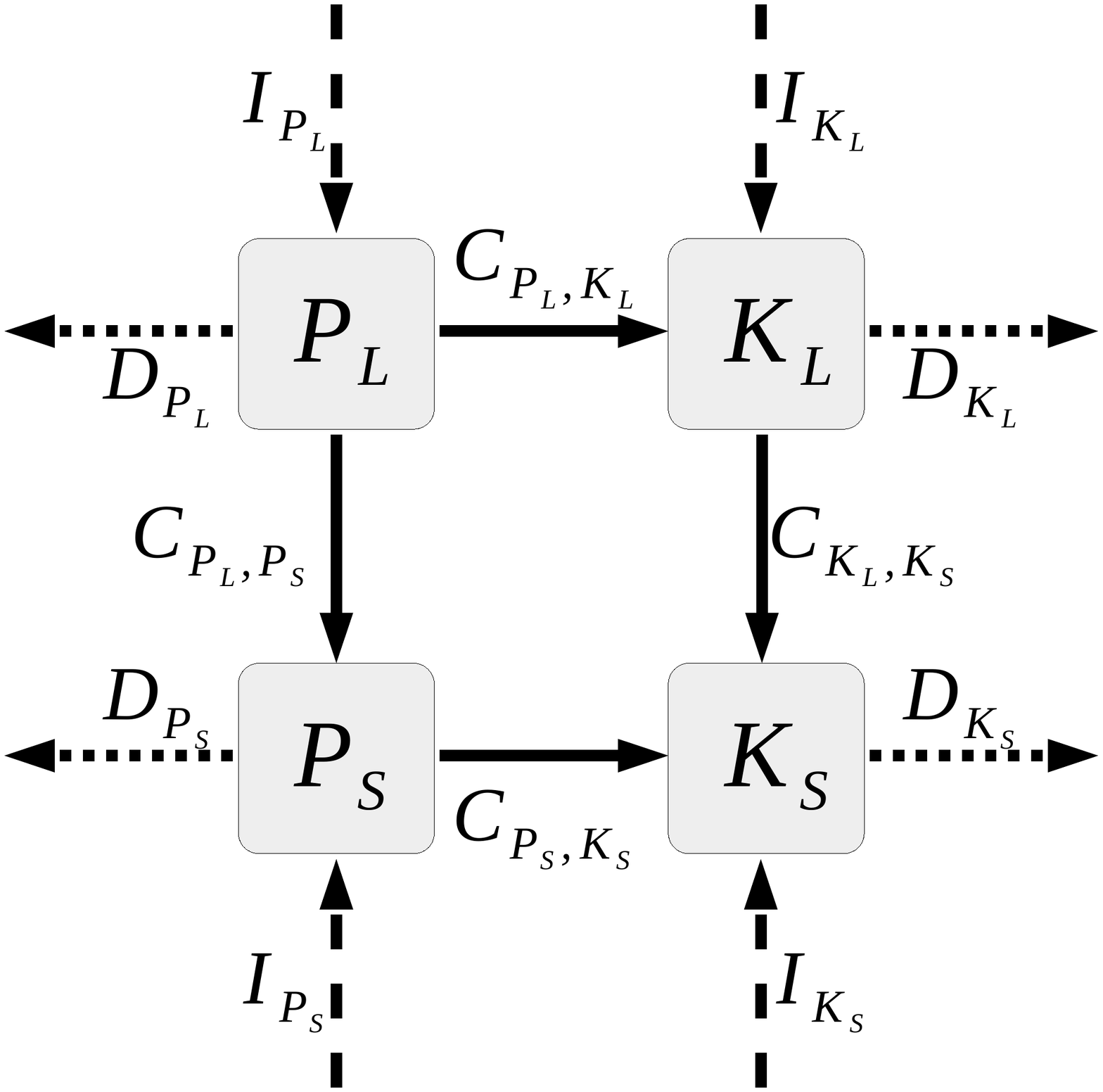}
\caption{\label{BnP_2-layer} Diagram of the Lorenz energy cycle of the two-level model. We indicate the fluxes of energy between large (L) and small (S) scales and between kinetic (K) and potential (P) energy. Dashed lines represent input of energy; dotted lines represent energy dissipation; and solid lines represent energy conversion terms.}
\end{figure}

Below we show the equations for the fluxes of energy for $X$ and $\theta$ alongside with those for $Y$ and $\phi$, as we have done in Section \ref{One-layer model} for the one-level model.

We define $K_L= \sum\limits_{k=1}^N {X_k^2}/{2}$ and $K_S= \sum\limits_{j=1}^J\sum\limits_{k=1}^N {Y_{j,k}^2}/{2}$ as the kinetic energy at  large and small scales, respectively. Similarly, we define $P_L= \sum\limits_{k=1}^N {\theta_k^2}/{2}$, and $P_S= \sum\limits_{j=1}^J\sum\limits_{k=1}^N {\phi_{j,k}^2}/{2}$, as the potential energy at  large and small scales, respectively. One derives the following equations for the various reservoirs of energy:
\begin{align}
	\frac{d}{dt}K_L&=I_{K_L}-C_{K_L,K_S}+C_{P_L,K_L}-D_{K_L}\label{EKL}\\
	\frac{d}{dt}K_S&=I_{K_S}+C_{K_L,K_S}+C_{P_S,K_S}-D_{K_S}\label{EKS}\\
	\frac{d}{dt}P_L&=I_{P_L}-C_{P_L,P_S}-C_{P_L,K_L}-D_{P_L}\label{EPL}\\
	\frac{d}{dt}P_S&=I_{P_S}+C_{P_L,P_S}-C_{P_S,K_S}-D_{P_S}\label{EPS}
\end{align}
where
\begin{equation}
\begin{split}	
I_{K_L}=F\sum\limits_{k=1}^N X_k\quad I_{K_S}=f\sum\limits_{j=1}^J\sum\limits_{k=1}^N Y_{j,k} \quad D_{K_L}=2\gamma K_L\quad D_{K_S}=2\gamma c K_S \\
I_{P_L}=G\sum\limits_{k=1}^N \theta_k\quad I_{P_S}=g\sum\limits_{j=1}^J\sum\limits_{k=1}^N \phi_{j,k} \quad D_{P_L}=2\gamma P_L\quad D_{P_S}=2\gamma c P_S\\
C_{K_L,K_S}=\frac{hc}{b}\sum\limits_{k=1}^N \sum\limits_{j=1}^J X_k Y_{j,k}\quad C_{P_L,P_S}=\frac{hc}{b}\sum\limits_{k=1}^N \sum\limits_{j=1}^J \theta_k \phi_{j,k}\\
C_{P_L,K_L}=-\alpha\sum\limits_{k=1}^N X_k \theta_k\quad C_{P_S,K_S}=-\alpha c \sum\limits_{k=1}^N \sum\limits_{j=1}^J Y_{j,k} \phi_{j,k}
\end{split}
\end{equation}
The energy cycle of this model is depicted in Fig. \ref{BnP_2-layer} and is closely reminiscent of the Lorenz energy cycle of the atmosphere.  Note that, if we add Eqs. \ref{EKL} with \ref{EKS}, on the one hand, and Eqs. \ref{EPL} with \ref{EPS}, on the other hand, we derive the energy budget for the total kinetic and total potential energy, respectively. Instead, if we add Eqs. \ref{EKL} with \ref{EPL}, on the one hand, and Eqs. \ref{EKS} with \ref{EPS}, on the other hand, we derive the budgets for the total energy at large and small scale, respectively.

\section*{Acknowledgements}
GV wishes to thank C. Franzke for several useful discussions. VL wishes to thank R. Blender and G. Gallavotti for the many inspiring conversations on the topics covered in this paper. VL acknowledges the support provided by the EU Horizon 2020 project TiPES (grant No. 820970) and by the EPSRC project "Applied Nonautonomous Dynamical Systems: Theory, Methods and Examples" (grant No. EP/T018178/1) The two authors have equally contributed to this paper. {\color{black}Scripts and data used for the preparation of this paper can be found at:\\\texttt{https://figshare.com/articles/journal\_contribution/VissioLucarini2020.zip/12917984}}.

%
%
%

%

%
%

\end{document}